\documentclass[fleqn,usenatbib]{mnras}
\usepackage{newtxtext,newtxmath}

\usepackage[T1]{fontenc}

\DeclareRobustCommand{\VAN}[3]{#2}
\let\VANthebibliography\thebibliography
\def\thebibliography{\DeclareRobustCommand{\VAN}[3]{##3}\VANthebibliography}

\usepackage{graphicx}	
\usepackage{amsmath}	
\usepackage{multirow}
\usepackage{tabularx}
\usepackage{gensymb}
\usepackage{caption}
\usepackage{subcaption}
\usepackage[T1]{fontenc}
\usepackage[dvipsnames]{xcolor}

\usepackage{xcolor}
\usepackage{hyperref}
\newcommand{\orc}{\includegraphics[height=\fontcharht\font`A]{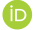}}
\newcommand{\orcid}[1]{\href{https://orcid.org/#1}{\orc}}

\newcolumntype{N}{>{\centering\arraybackslash}m{0.95in}}
\newcolumntype{G}{>{\centering\arraybackslash}m{0.75in}}
\newcolumntype{F}{>{\centering\arraybackslash}m{1.14in}}
\newcolumntype{K}{>{\centering\arraybackslash}m{1.06in}}
\newcolumntype{J}{>{\centering\arraybackslash}m{1.18in}}

\title[NGTS-23b, 24b, 25b and HATS-54b from NGTS]{The discovery of three hot Jupiters, NGTS-23b, 24b and 25b, and updated parameters for HATS-54b from the Next Generation Transit Survey}

\author[David G. Jackson et al.]{David G. Jackson\orcid{0000-0003-0343-7905}$^{1}$\thanks{E-mail: djackson823@qub.ac.uk},
Christopher A. Watson$^{1}$,
Ernst J. W. de Mooij$^{1}$,
Jack S. Acton$^2$,
Douglas R. Alves$^{3}$,
\newauthor  David R. Anderson,$^{4,5}$
David J. Armstrong$^{4,5}$,
Daniel Bayliss$^{4,5}$,
Claudia Belardi$^{2}$,
Fran\c{c}ois Bouchy$^{6}$,
\newauthor Edward M. Bryant$^{7}$,
Matthew R. Burleigh$^{2}$,
Sarah L. Casewell$^2$,
Jean C. Costes$^{8}$,
Phillip Eigm\"uller$^9$,
\newauthor Michael R. Goad$^2$,
Samuel Gill$^{4,5}$,
Edward Gillen,$^{10,11}$,
Maximilian N. Günther$^{12,13}$,
Faith Hawthorn$^{4,5}$,
\newauthor Beth A. Henderson$^2$,
James A. G. Jackman$^{14,4,5}$,
James S. Jenkins$^{15,16}$,
Monika Lendl$^{6}$, 
Alicia Kendall$^2$,
\newauthor James McCormac$^{4,5}$,
{Maximiliano Moyano}$^{17}$,
Louise. D. Nielsen$^{18}$,
Ares Osborn$^{4,5}$,
Ramotholo~R.~Sefako$^{19}$, 
\newauthor Alexis M. S. Smith$^{9}$,
Rosanna H. Tilbrook$^2$,
Oliver Turner$^{6}$,
Stéphane Udry$^{6}$, 
Jose I. Vines$^{3}$, 
Richard G. West$^{4,5}$,
\newauthor Peter J. Wheatley$^{4,5}$
and
Hannah Worters$^{19}$
\\
$^{1}$Astrophysics Research Centre, School of Mathematics and Physics, Queen’s University Belfast, BT7 1NN Belfast, UK \\
$^{2}$School of Physics and Astronomy, University of Leicester, University Road, Leicester, LE1 7RH, UK\\
$^{3}$Departamento de Astronomia, Universidad de Chile, Casilla 36-D, Santiago, Chile\\
$^{4}$Centre for Exoplanets and Habitability, University of Warwick, Gibbet Hill Road, Coventry, CV4 7AL, UK\\
$^{5}$Dept. of Physics, University of Warwick, Gibbet Hill Road, Coventry, CV4 7AL, UK\\
$^{6}$Observatoire de Gen\`eve, Universit\'e de Gen\`eve, Chemin Pegasi 51, 1290 Sauverny, Switzerland\\
$^{7}$Mullard Space Science Laboratory, University College London, Holmbury St Mary, Dorking, Surrey, RH5 6NT, UK\\
$^{8}$Aix Marseille Univ, CNRS, CNES, LAM, Marseille, France\\
$^{9}$Institute of Planetary Research, German Aerospace Center, Rutherfordstrasse 2, 12489 Berlin, Germany\\
$^{10}$Astronomy Unit, Queen Mary University of London, Mile End Road, London E1 4NS, UK\\
$^{11}$Astrophysics Group, Cavendish Laboratory, J.J. Thomson Avenue, Cambridge, CB3 0HE, UK\\
$^{12}$European Space Agency (ESA), European Space Research and Technology Centre (ESTEC), Keplerlaan 1, 2201 AZ Noordwijk, Netherlands\\
$^{13}$ESA Research Fellow\\
$^{14}$School of Earth and Space Exploration, Arizona State University, Tempe, AZ, 85287, USA\\
$^{15}$N\'ucleo de Astronom\'ia, Facultad de Ingenier\'ia y Ciencias, Universidad Diego Portales, Av. Ej\'ercito 441, Santiago, Chile\\
$^{16}$Centro de Astrofísica y Tecnologías Afines (CATA), Casilla 36-D, Santiago, Chile\\
$^{17}$ {Instituto de Astronom\'ia, Universidad Cat\'olica del Norte,
Angamos 0610, 1270709, Antofagasta, Chile}\\
$^{18}$European Southern Observatory (ESO), Karl-Schwarzschild-Stra{\ss}e 2, 85748 Garching bei M{\"u}nchen, Germany\\
$^{19}$South African Astronomical Observatory, P.O Box 9, Observatory 7935, Cape Town, South Africa\\
}

\date{Accepted XXX. Received YYY; in original form ZZZ}

\pubyear{2022}

\begin{document}
\label{firstpage}
\pagerange{\pageref{firstpage}--\pageref{lastpage}}
\maketitle

\begin{abstract}
We report the discovery of three new hot Jupiters with the Next Generation Transit Survey (NGTS) as well as updated parameters for HATS-54b, which was independently discovered by NGTS. NGTS-23b, NGTS-24b and NGTS-25b have orbital periods of 4.076, 3.468, and 2.823 days and orbit G-, F- and K-type stars, respectively. NGTS-24 and HATS-54 appear close to transitioning off the main-sequence (if they are not already doing so), and therefore are interesting targets given the observed lack of hot Jupiters around sub-giant stars. By considering the host star luminosities and the planets’ small orbital separations (0.037 -- 0.050 au), we find that all four hot Jupiters are above the minimum irradiance threshold for inflation mechanisms to be effective. NGTS-23b has a mass of 0.61\,$M_{\rm{J}}$ and radius of 1.27\,$R_{\rm J}$ and is likely inflated. With a radius of 1.21\,$R_{\rm J}$ and mass of 0.52\,$M_{\rm J}$, NGTS-24b has a radius larger than expected from non-inflated models but its radius is smaller than the predicted radius from current Bayesian inflationary models. Finally, NGTS-25b is intermediate between the inflated and non-inflated cases, having a mass of 0.64\,$M_{\rm J}$ and a radius of 1.02\,$R_{\rm J}$. The physical processes driving radius inflation remain poorly understood, and by building the sample of hot Jupiters we can aim to identify the additional controlling parameters, such as metallicity and stellar age.

\end{abstract}

\begin{keywords}
planetary systems -- planets and satellites: detection -- stars: individual: NGTS-23 (CTOI-77287067), NGTS-24 (TOI-4270), NGTS-25 and HATS-54
\end{keywords}



\section{Introduction}\label{Sec:Intro}

Although hot Jupiters have a low occurrence rate of <1\% \citep{Mayor2011, Hsu2019, Zhou2019}, they still represent one of the most common categories of planets found due to biases inherent in exoplanet discovery techniques. Understanding the structure, composition and evolution of these planets is of fundamental importance in understanding the processes of formation, evolution and migration acting on the wider exoplanet population. Increasing the sample of known hot Jupiters continues to help constrain and parameterize planetary models. For example, there are models that attempt to explain the evolution of planetary radii whilst accounting for the effects of stellar irradiation and heavy element cores \citep{Fortney2007,Baraffe2008} but they struggle to explain the observed radii of highly-irradiated gas giants.

There are a number of proposed mechanisms to explain inflation in hot Jupiters such as kinetic heating \citep{Guillot2002}, Ohmic heating through magnetohydrodynamic effects \citep{Batygin2010, Perna2010, Ginzburg2016}, double diffusive convection \citep{Chabrier2007} and tidal dissipation \citep{Bodenheimer2001, Bodenheimer2003, Jermyn2017}. In order to distinguish between such models, a robust sample of well-characterised hot Jupiters is required to enable statistical studies to constrain these models (such as that by \citealt{Sestovic2018}). Of particular interest are planets irradiated by fluxes greater than approximately $2\times10^5$ $\rm{W\,m^{-2}}$, which appear to be preferentially inflated \citep{Demory2011, Miller2011}. However, it is still unclear what impact additional parameters, such as metallicity, have on whether the planet will appear inflated.\

We present the discovery of three new hot Jupiters from the Next Generation Transit Survey \citep[NGTS;][]{Wheatley2018}. We also refine the parameters for HATS-54b, which was independently discovered by NGTS. In Section \ref{Sec:Obs}, the initial NGTS photometry is described as well as the follow-up observations (both photometry and spectroscopy). Section \ref{Sec:Analysis} describes how both the stellar and planetary parameters were derived. The results will be discussed in Section \ref{Sec:Results} including investigations into the orbital decay timescales (Section \ref{sec:decay}) and the potential inflation (Section \ref{sec:inflation}) of these planets. Our final conclusions are given in Section \ref{Sec:Conclusions}.

\section{Observations}\label{Sec:Obs}

The planets presented in this work were confirmed using both photometric and spectroscopic observations. These observations are outlined in this section.
 
\subsection{Photometry}
All planets presented in this paper were detected using photometry from NGTS. To confirm the planetary nature of the stellar companions, and to further constrain the system parameters, additional photometry was obtained using the Transiting Exoplanet Survey Satellite \citep[TESS;][]{Ricker2014}, EulerCam \citep{Lendl2012} and the 1-/1.9\,m telescopes at the South African Astronomical Observatory \citep[SAAO;][]{Coppejans2013}. A summary of the photometric observations is given in Table \ref{tbl:Phot Sum}, and the phase-folded transit lightcurves can be seen in Figure \ref{fig:photmetry plots}. Lightcurves not used in the global fitting can be found in the appendix (Figure \ref{fig:appendix phot plots}).

\begin{table*}
\caption{Summary of photometric observations.}
\begin{center}
 \begin{tabular}{c c c c c c c}
 \hline
 Target & Instrument & Nights Observed & N$_{\rm{images}}$ & Exptime (s) & Filter & Comments\\ [0.5ex] 
 \hline\hline
 \multirow{8}{5em}{NGTS-23} & NGTS & 2017/08/16 -- 2018/03/18 & 187,478 & 10 & NGTS &\\ 
 & TESS & 2018/10/19 -- 2018/11/14 & 1027 & 1800 & TESS & Sector 4 \\
 & TESS & 2018/11/15 -- 2018/12/11 & 1176 & 1800 & TESS & Sector 5\\
 & TESS & 2020/10/22 -- 2020/11/16 & 3445 & 600 & TESS & Sector 31\\
 & TESS & 2020/11/20 -- 2020/12/16 & 3589 & 600 & TESS & Sector 32\\
 & EulerCam & 2018/12/27 & 171 & 90 & VG &\\
 & SAAO 1.0\,m & 2018/12/15 & 300 & 60 & V &\\
 & SAAO 1.9\,m & 2018/12/19 & 295 & 60 & V &\\
 \hline
 \multirow{4}{5em}{NGTS-24} & NGTS & 2017/12/10 -- 2018/06/26 & 235,369 & 10 & NGTS &\\ 
 & TESS & 2019/02/28 -- 2019/03/25 & 1084 & 1800 & TESS & Sector 9\\
 & TESS & 2019/03/26 -- 2019/04/22 & 1103 & 1800 & TESS & Sector 10\\
 & TESS & 2021/03/07 -- 2021/04/01 & 3467 & 600 & TESS & Sector 36\\
 \hline
  \multirow{3}{5em}{NGTS-25} & NGTS & 2018/03/24 -- 2018/11/01 & 208,001 & 10 & NGTS &\\ 
 & TESS & 2020/07/05 -- 2020/07/30 & 2898 & 600 & TESS & Sector 27\\
 & SAAO 1.0\,m & 2019/10/03 & 940 & 20 & V &\\
 \hline
 \multirow{7}{5em}{HATS-54\\(NGTS-22)} & NGTS & 2015/12/18 -- 2016/09/01 & 124,203 & 10 & NGTS &\\ 
 & TESS & 2019/04/23 -- 2019/05/20 & 1145 & 1800 & TESS & Sector 11\\
 & TESS & 2021/04/02 -- 2021/04/28 & 3461 & 600 & TESS & Sector 37\\
 & EulerCam & 2019/03/25 & 130 & $\sim$80 & VG &\\
 & EulerCam & 2019/04/27 & 136 & 75 & VG &\\
 & SAAO 1.0\,m & 2018/07/09 & 700 & 20 & I &\\
 & SAAO 1.0\,m & 2019/04/30 & 360 & 60 & V &\\
 \hline

\end{tabular}
\label{tbl:Phot Sum}
\end{center}
\end{table*}

\subsubsection{NGTS}

NGTS is a wide-field photometric survey located at ESO's Paranal Observatory in Chile. NGTS has a fully robotic array of twelve 20cm Newtonian telescopes operating using the unique NGTS filter band (520-890nm). Each telescope is equipped with a 2K$\times$2K e2V deep-depleted Andor IKon-L CCD camera. It is optimised to detect and characterise transiting exoplanets around K and early M-type stars and has achieved sufficient precision to discover exoplanets as small as 3$R_{\oplus}$  \citep{Wheatley2018, West2019, Smith2021}.\

NGTS-23 (CTOI-77287067; {V=14.010}), NGTS-24 (TOI-4270; {V=13.192}) and NGTS-25({V=14.266}) were observed with NGTS during the 2018 campaign for a total of 162 nights, 158 nights and 147 nights, respectively. HATS-54 (TOI-1920, NGTS-22; {V=13.914}) was observed during the 2016 campaign for a total of 119 nights and was detected by NGTS independently of its discovery by  \citet{Espinoza2019}. All the objects were observed with a 10s exposure time and over 120,000 images were collected for each object. The data were reduced and aperture photometry performed via the CASUTools \footnote{\url{http://casu.ast.cam.ac.uk/surveys-projects/software-release}} package, before nightly trends were removed using an adapted version of the \textsc{SysRem} algorithm \citep{Tamuz2005}.\

The data was then searched for transit-like events by \textsc{ORION}, a custom implementation of the box-fitting least-squares (BLS) fitting algorithm \citep{Kovacs2002, CollierCameron2006, Wheatley2018}. \textsc{ORION} identified nine partial and one full transit for HATS-54b (NGTS-22b);  ten partial and one full transit for NGTS-23b; nine partial and seven full transits for NGTS-24b; and eleven partial and eight full transits for NGTS-25b. The initial fits to the NGTS data provided by \textsc{ORION} revealed that the depths, widths and shapes of the transits for each object were compatible with transiting hot Jupiters. \

Further checks were performed to ensure that the nature of the signal was planetary and not stellar or a false positive. These included comparing the transit depth of consecutive odd and even transits, which can show a depth difference that can demonstrate an eclipsing binary system has been mis-folded on half the true period. Also, secondary eclipses around 0.5 phase in the phase-folded lightcurves were searched for as this could also indicate that the orbiting companion is stellar in nature.  All candidates passed this vetting and thus follow-up observations were taken.\

\subsubsection{SAAO}\label{sec:SAAO}

Between 2018 July and 2019 October, we took follow-up photometry for each target (except for NGTS-24) using either the 1.0\,m or 1.9\,m telescope at SAAO \citep{Coppejans2013}. The 1.0\,m telescope was equipped with the full-frame transfer CCD Sutherland High-speed Optical Camera (SHOC), “SHOC’n’awe", that has a 2.85\arcmin $\times$2.85\arcmin field of view. The 1.9\,m telescope was equipped with the “SHOC’n’disbelief” optical camera, that had a slightly narrower field of view of 1.3\arcmin $\times$1.3\arcmin. {We were able to simultaneously observe at least one comparison star of similar brightness for each object.} All observations were taken using either the I band or V band filters. \

Two lightcurves were obtained for NGTS-23 one each on the 1.0\,m and 1.9\,m telescopes, on the 2018 December 15 and 2018 December 19, respectively. Both observations used an exposure time of 60 seconds with the V band filter. Both observations ended early due to high humidity however, we obtained 300 images and 295 images, respectively.\

A single V-band light curve was taken for NGTS-25b on the 2019 October 3 using the 1.0\,m telescope. 940 images were obtained using an exposure time of 20 seconds. However, the middle of these observations were affected by bad seeing. \

Two observations of HATS-54b were conducted using the 1.0\,m telescope on the nights of the 2018 July 9 and the 2019 April 30. The 2018 observation obtained 700 images with an exposure time of 20s using the I band filter. This observation was ended early due to high humidity. The 2019 observation obtained 360 images with an exposure time of 60s using the V band filter.\

Each lightcurve was bias and flat-field corrected using the local Python-based SAAO SHOC pipeline, which uses \textsc{IRAF} \citep{Tody1986} photometry tasks \citep[\textsc{PyRAF}-][]{PyRAF2012} and facilitates the extraction of raw and differential lightcurves. We used the Starlink package \textsc{autophotom} \citep{Currie2014} to perform aperture photometry on both our target and comparison stars, and chose apertures that gave the maximum signal-to-noise ratio. Background apertures were adjusted to account for changes in the stellar point-spread function over the night as the atmospheric conditions varied. Finally, the measured fluxes of the comparison stars for each object were used for differential photometry of our targets.\

\subsubsection{EulerCam}

We also observed NGTS-23 and HATS-54 with EulerCam \citep{Lendl2012} on the 1.2 m Euler Telescope at La Silla Observatory. All observations presented in this work were done through a \emph{Geneva V} filter \citep{Rufener1988}. NGTS-23 was observed on the 2018 December 27, with 171 images being acquired with an exposure time of 90\,s and a 0.1\,mm defocus. \

HATS-54 was observed twice with EulerCam. The first observation was performed on the 2019 March 5, with 130 images being acquired with an exposure time of approximately 80 seconds. No defocus was applied for this observation. The second observation was carried out on the 2019 April 27. 136 images were acquired with an exposure time of 75 seconds and a defocus of 0.5\,mm. Only a partial transit of NGTS-23 was observed, whereas the observations of HATS-54b covered one full transit and one partial transit.\

For both targets, the data were reduced using the standard procedure of bias subtraction and flat field correction. The aperture photometry as well as x- and y-positions, full width half-maximum (FWHM), airmass and sky background of the target star were extracted using the PyRAF-based routines. The comparison stars and the photometric aperture radius were carefully chosen in order to reduce the RMS in the out-of-transit lightcurve parts.\

\subsubsection{TESS}

TESS is a space-based NASA survey telescope that searches for transiting planets around bright stars \citep{Ricker2014}. It has a wide field of view, with four 24$\times$24\textdegree~ cameras, each equipped with four 2k$\times$2k CCDs. Its typical observing baseline of $\approx$27 days makes it well-suited to detecting short-period transiting exoplanets. We searched for our candidates in the TESS full frame images (FFIs) using the TESSCut \footnote{\url{https://mast.stsci.edu/tesscut/}} tool and found that all four stars were observed by TESS, with observations between the 2018 October 19 and the 2021 April 28. NGTS-23 was observed in Sectors 4, 5, 31 and 32; NGTS-24 was observed in Sectors 9, 10 and 36; HATS-54 was observed in Sectors 11 and 37 and NGTS-25 was observed in Sector 27. HATS-54b falls onto a CCD in TESS Sector 38 but we do not use it as it appears to be in the overscan region of the camera. A summary of this information can be found in Table \ref{tbl:Phot Sum}.\

Observations before 2020 July were within the primary mission of TESS, so the TESS FFIs for these sectors were observed at 30 minute cadence. For observations that are in the first extended mission of TESS (which is set to finish in 2022 September) the TESS FFIs were observed at 10 minute cadence. This means that all of the stars were observed at both 10 minute and 30 minute cadence with TESS apart from NGTS-25, which was only observed with 10 minute cadence. For each of our candidates we used an automatically determined optimal aperture to exclude neighbouring stars, ranging from three to eight pixels. However, the large pixel scale of TESS (21\arcsec) meant that there was still some slight dilution of the FFI lightcurves.\

For each candidate, we found that the best BLS period from TESS was consistent with the \textsc{ORION} value for the NGTS photometry. TESS detected six full transits for NGTS-25, eighteen full transits for NGTS-24, two partial and twenty full transits for NGTS-23 and seventeen full transits for HATS-54. \

For all four objects, the transits observed with TESS, as well as the other photometric follow-up, was consistent with planetary companions. Therefore, all objects were subsequently followed up with spectroscopy to determine the mass of the companions.

\begin{figure*}
    \centering
    \begin{subfigure}[b]{0.475\textwidth}
        \centering
        \includegraphics[width=\textwidth]{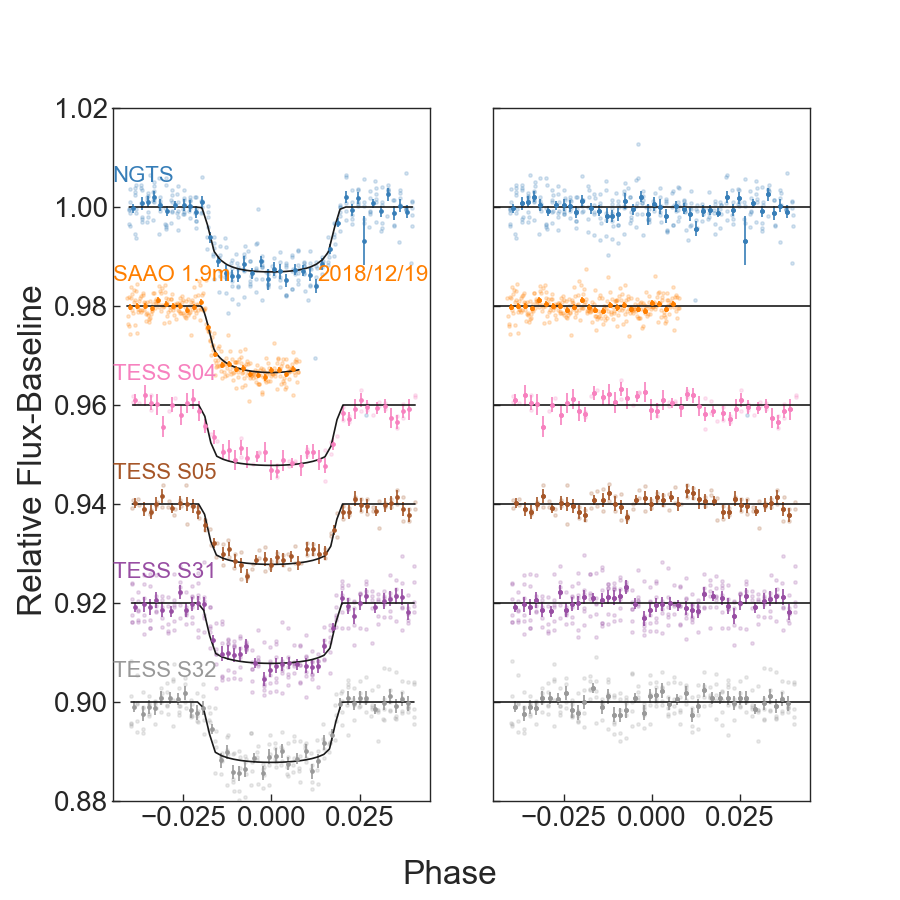}
        \caption[]%
        {{\small NGTS-23}}    
        \label{fig:photometry NGTS-23}
    \end{subfigure}
    \hfill
    \begin{subfigure}[b]{0.475\textwidth}  
        \centering 
        \includegraphics[width=\textwidth]{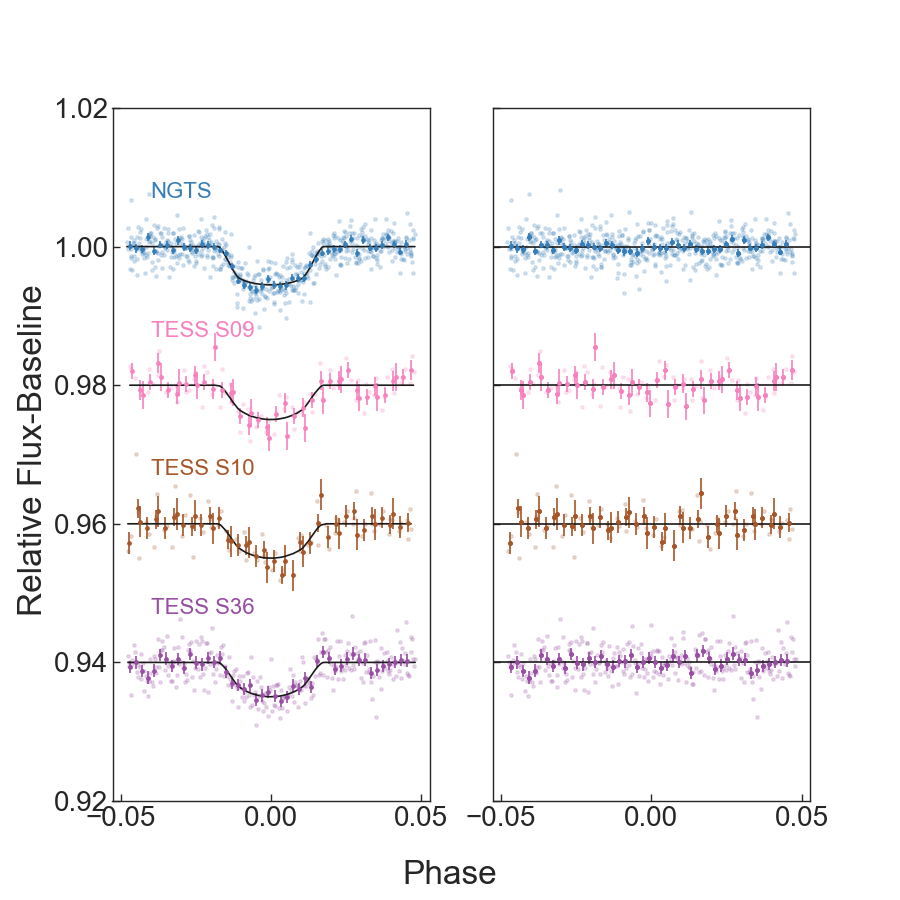}
        \caption[]%
        {{\small NGTS-24}}    
        \label{fig:photometry NGTS-24}
    \end{subfigure}
    \vskip\baselineskip
        \begin{subfigure}[b]{0.475\textwidth}   
        \centering 
        \includegraphics[width=\textwidth]{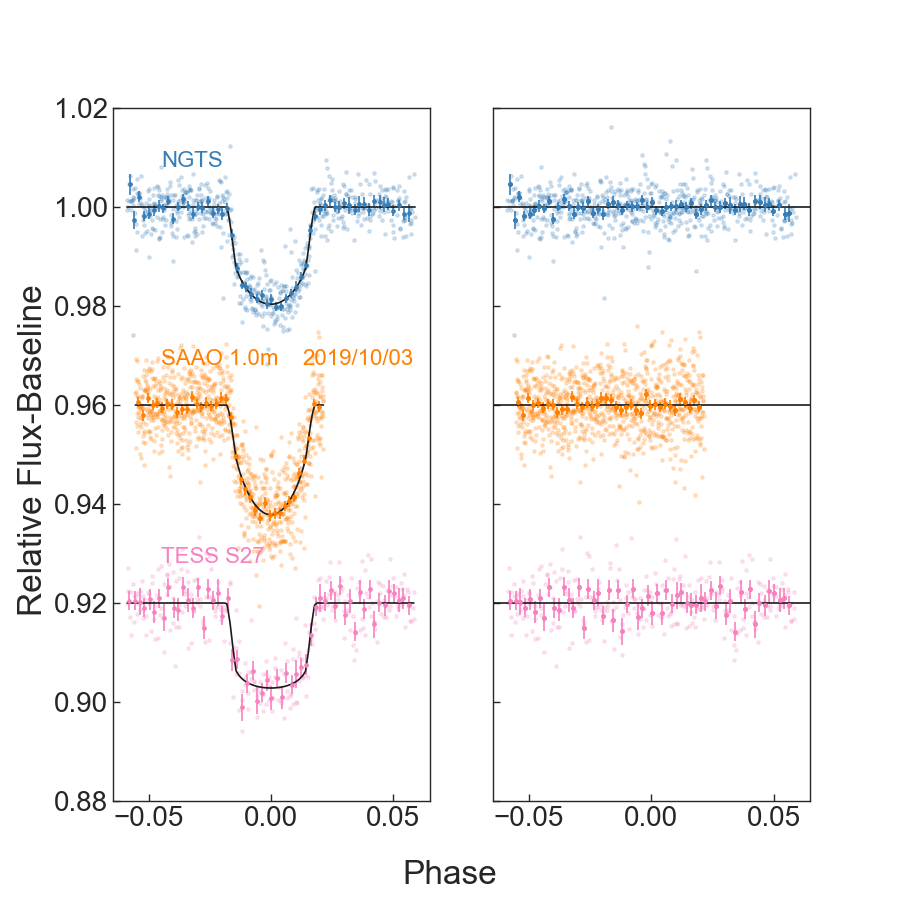}
        \caption[]%
        {{\small NGTS-25}}    
        \label{fig:photometry NGTS-25}
    \end{subfigure}
    \hfill
    \begin{subfigure}[b]{0.475\textwidth}   
        \centering 
        \includegraphics[width=\textwidth]{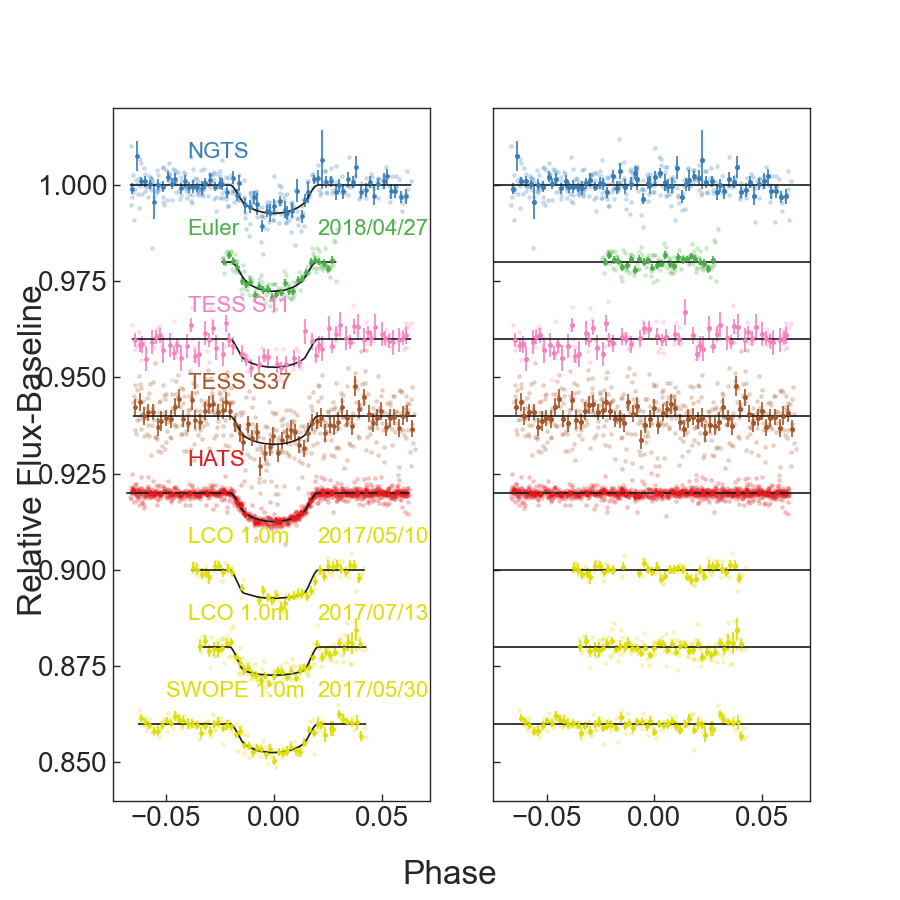}
        \caption[]%
        {{\small HATS-54 (NGTS-22)}}    
        \label{fig:photometry HATS-54b}
    \end{subfigure}
    
    \caption[]
    {\small NGTS discovery photometry (blue points) and follow up photometry for each planet. The lighter points show the original photometry except for NGTS, which has been binned to 10 minute intervals. The darker datapoints have been binned to 0.002 phase. Photometry from HATS is shown in red while the HATS follow-up photometry that was also used in our combined fits are shown in yellow. The solid black line shows the lightcurve model for the planet. Residuals are shown to the right of the lightcurves.  Lightcurves offset for clarity.} 
    \label{fig:photmetry plots}
\end{figure*}

\subsection{Spectroscopy}

\begin{figure*}
    \centering
    \begin{subfigure}[b]{0.49\textwidth}
        \centering
        \includegraphics[width=\textwidth]{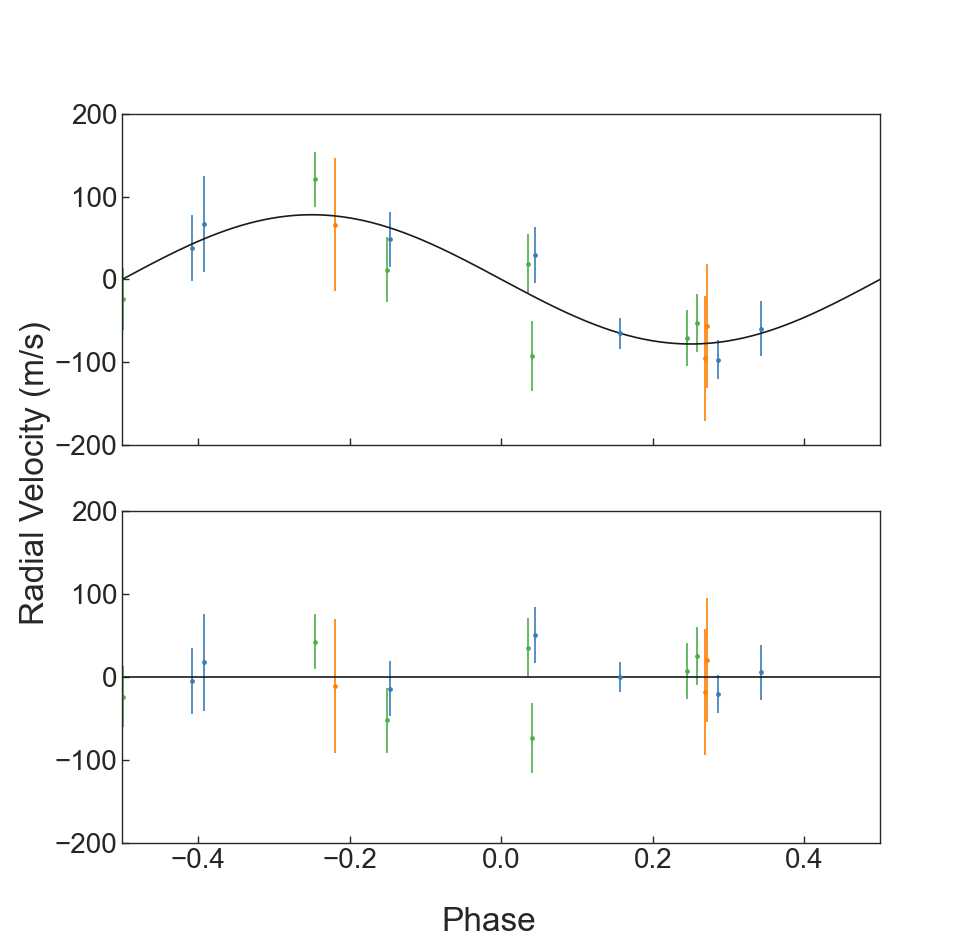}
        \caption[]%
        {{\small NGTS-23}}    
        \label{fig:rv NGTS-23b}
    \end{subfigure}
    \hfill
    \begin{subfigure}[b]{0.49\textwidth}  
        \centering 
        \includegraphics[width=\textwidth]{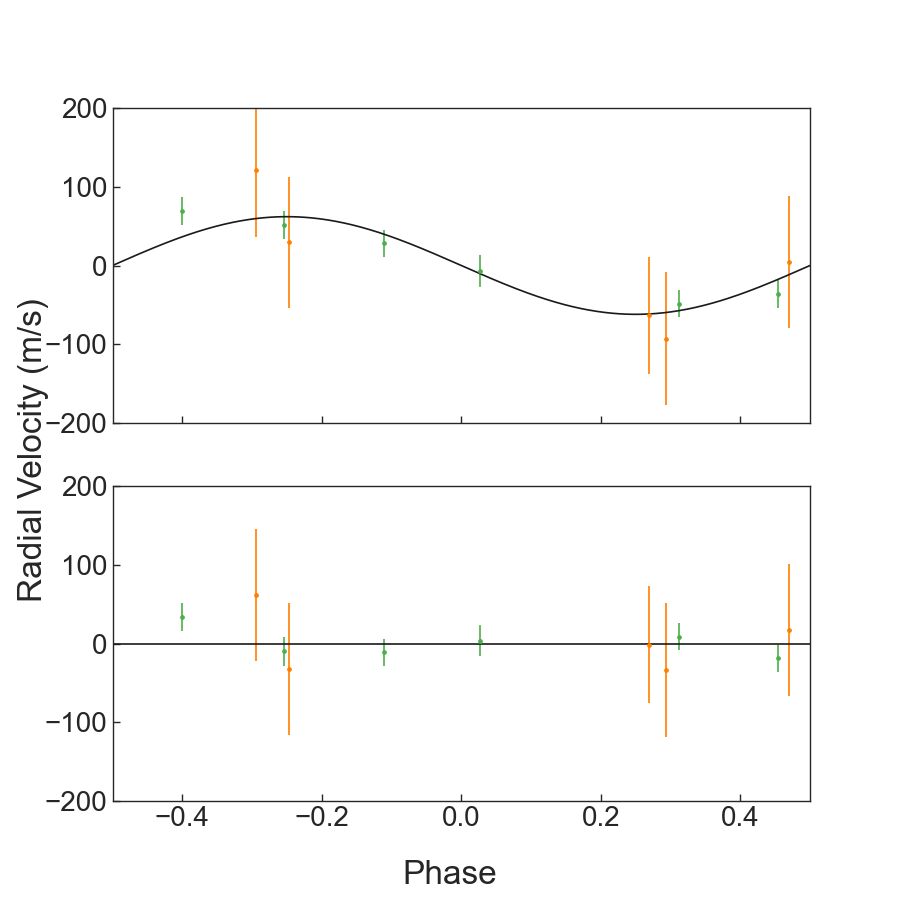}
        \caption[]%
        {{\small NGTS-24}}    
        \label{fig:rv NGTS-24b}
    \end{subfigure}
    \vskip\baselineskip
        \begin{subfigure}[b]{0.49\textwidth}   
        \centering 
        \includegraphics[width=\textwidth]{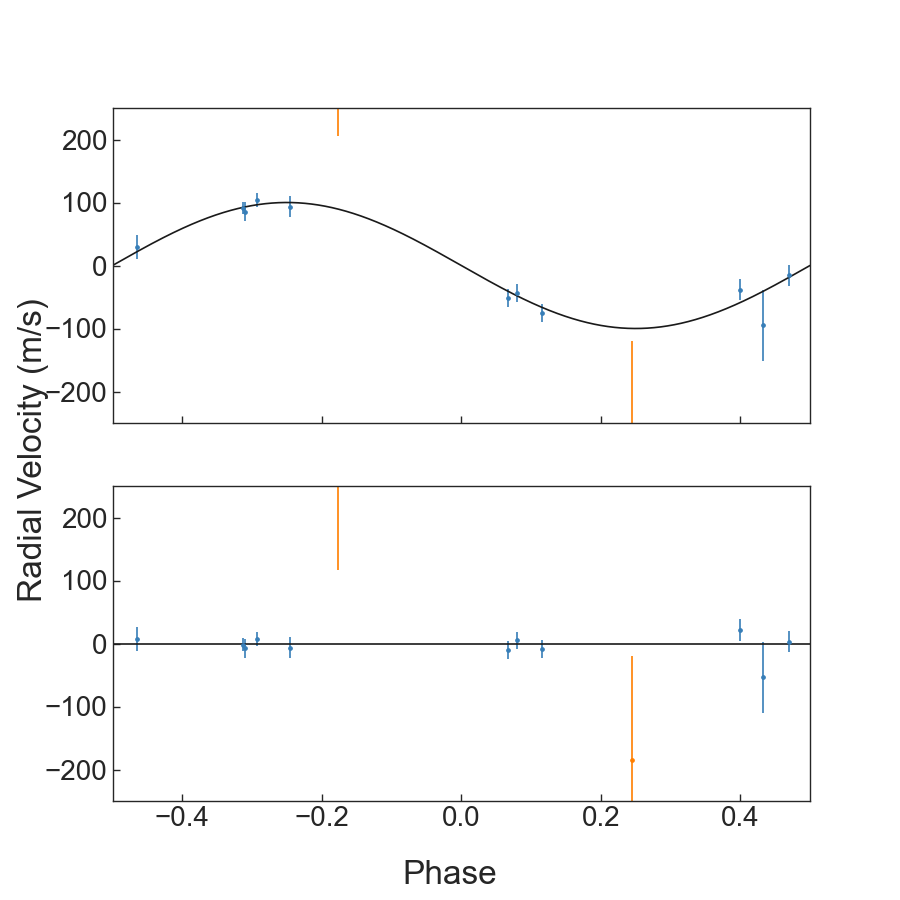}
        \caption[]%
        {{\small NGTS-25}}    
        \label{fig:rv NGTS-25b}
    \end{subfigure}
    \hfill
    \begin{subfigure}[b]{0.49\textwidth}   
        \centering 
        \includegraphics[width=\textwidth]{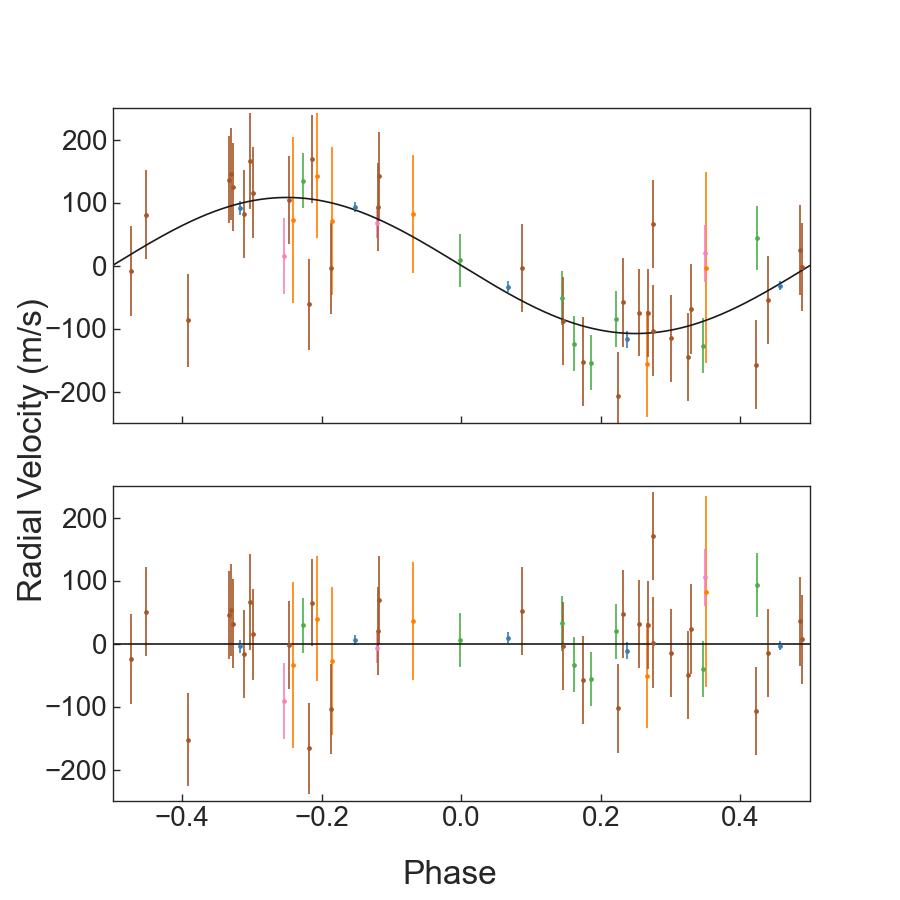}
        \caption[]%
        {{\small HATS-54 (NGTS-22)}}    
        \label{fig:rv HATS-54b}
    \end{subfigure}
    \caption[]
    {\small Phase-folded radial velocity measurements for NGTS-23, NGTS-24, NGTS-25, and HATS-54 (NGTS-22). These include data obtained from HARPS (blue points), FEROS (green points) and CORALIE (orange points). HARPS spectroscopy used by the HATS team is shown in peak while the FEROS spectroscopy used by HATS is shown in brown. The radial velocity model is shown by the solid black line. The residuals are shown below the radial velocity measurements.} 
    \label{fig:rv plots}
\end{figure*}

In addition to the photometric follow-up, spectroscopy was also used to obtain radial velocity measurements for all of the systems. This allowed for the mass, and therefore the density, of the respective transiting companion to be determined. A summary of the spectroscopic observations is given in Table \ref{tbl:Spec Sum} and phase-folded radial-velocity curves can be seen in Figure \ref{fig:rv plots}.\ 

The three different \'echelle spectrographs used are all located at La Silla Observatory in Chile. The High Accuracy Radial velocity Planet Searcher \citep[HARPS;][]{Mayor2003} has a resolving power of R=115,000 and is mounted on the ESO 3.6\,m telescope. The R=60,000 CORALIE spectrograph \citep{Queloz2001} is installed on the Swiss 1.2\,m Leonhard Euler telescope while the Fibre-fed Extended Range Optical Spectrograph \citep[FEROS;][]{Kaufer1998} is mounted on the MPG/ESO 2.2\,m telescope and has R=48,000. All four systems were observed by a minimum of two spectrographs and have at least 11 datapoints (see Table \ref{tbl:Spec Sum}).\

The HARPS data was reduced via the offline data reduction system (DRS) before being cross-correlated with a binary mask to extract radial velocities \citep{Baranne1996}. For NGTS-24 and NGTS-23 a G2 mask was used whereas a G9 mask and a K0 mask were used for HATS-54 and NGTS-25, respectively. This procedure was also employed for the CORALIE data, for which the spectra were reduced using the standard CORALIE DRS. For the data collected with FEROS, the CERES reduction pipeline \citep{Brahm2017} performed a radial velocity extraction by the same method.\

All four targets had semi-amplitudes consistent with a Jupiter-mass companion ($K$ $\sim$ 100\,$\rm{m\,s^{-1}}$) and had variations in phase with the derived period from the photometric data. We investigated whether there were any correlations between the line bisector (BIS) of the CCF to ensure that the signals were of planetary origin and did not arise from stellar activity (e.g. stellar spots) or a blended eclipsing binary \citep{Queloz2001b}. No evidence was found to contradict the existence of a planetary companion.

\begin{table*}
\caption{Summary of the spectroscopic observations. $\mathrm{SNR_{\rm{combined}}}$ is the SNR of the co-added spectra that were used in the analysis outlined in Section \ref{Sec:Analysis}.}
\begin{center}
 \begin{tabular}{c c c c c c  }
 \hline
 Target & Instrument & Nights Observed & N$_{\mathrm{spectra}}$ & $\mathrm{SNR}_{\mathrm{combined}}$  & {Programme}\\ [0.5ex] 
 \hline\hline
 \multirow{4}{*}{NGTS-23} & HARPS & 2019/09/12 -- 2020/01/20 & 7 & 29.13 & {0104.C-0588} \\ 
 & \multirow{2}{*}{FEROS} & \multirow{2}{*}{2018/12/24 -- 2020/01/04} & \multirow{2}{*}{8} & \multirow{2}{*}{-} & {0102.A-9011 \& 0103.A-9004}\\
 & & & & & {\& 0103.C-0719 \& 0104.A-9012} \\
 & CORALIE & 2018/10/27 -- 2018/11/06 & 3 & - & {N/A} \\
 \hline
 \multirow{2}{*}{NGTS-24} & FEROS & 2019/12/30 -- 2020/01/04 & 6 & - & {0104.A-9012} \\ 
 & CORALIE & 2019/04/22 -- 2019/12/26 & 5 & - & {N/A}\\
 \hline
  \multirow{2}{*}{NGTS-25} & HARPS & 2021/06/12 -- 2021/09/13 & 12 & 20.92 & {105.20G9}\\
 & CORALIE & 2019/08/09 -- 2021/08/22 & 2 & - & {N/A}\\
 \hline
 \multirow{3}{5em}{HATS-54\\ (NGTS-22)} & HARPS & 2019/05/10 -- 2019/06/08 & 5 & 26.45 & {0103.C-0719}\\ 
 & FEROS & 2019/04/18 -- 2019/04/26 & 8 & - & {0103.A-9004} \\
 & CORALIE & 2019/02/23 -- 2019/04/28 & 6 & - & {N/A}\\
 \hline
\end{tabular}
\label{tbl:Spec Sum}
\end{center}
\end{table*}

\section{Analysis}\label{Sec:Analysis}

The measured planetary parameters are dependent on 
the characterisation of 
the host star. Therefore, accurately constraining the properties of the host star is vital to determine accurate planetary parameters.

\subsection{Stellar Analysis}\label{sec:spec analysis}

{Most of our stellar parameters were obtained by fitting the spectral energy distributions (SEDs). However, to obtain suitable priors to constrain the SED fits, we first analysed spectra for each object. For the spectral analysis we used the open source framework \textsc{iSpec} \citep{Blanco-Cuaresma2014, Blanco-Cuaresma2019}. \textsc{iSpec} allows users to choose from a list of models for spectral synthesis. We chose to employ the \textsc{spectrum} \citep{Gray1994} radiative transfer code and the solar abundance model from \citet{Asplund2009}. Additionally, we use the Gaia-ESO Survey (GES) line list \citep[version 6.0 -][]{Heiter2021}, which covers most of the wavelength range of our spectra (420-920nm), and the MARCES.GES model atmosphere \citep{Gustafsson2008}.}\

We used our high-resolution HARPS data for the analysis of NGTS-23, NGTS-25 and HATS-54. For each object, we shift the spectra to the laboratory frame of reference before co-adding them to produce a high signal-to-noise ratio (SNR) combined spectrum. The SNRs of these spectra are reported in Table \ref{tbl:Spec Sum}. For NGTS-24, we attempted to follow a similar procedure using the CORALIE/FEROS spectra. However, the SNR of the co-added CORALIE spectra were too low to obtain meaningful results. For FEROS, the Th-Ar lamp used for calibration appears to have contaminated the stellar spectrum. Therefore, we are unable to use the FEROS spectra for spectral analysis but they still yield reliable results for the radial  velocity measurements.  \

We determined the stellar effective temperature, (${T_{\rm {eff}}}$) and the surface gravity (${log\,g}$), via fits to the the H$\alpha$, NaI D and MgI b lines whereas FeI and FeII lines were used to determine the metallicity (${[Fe/H]}$), alpha enhancement ($[\alpha/Fe]$), microscopic velocity (${v_{\rm{mic}}}$), macroscopic velocity (${v_{\rm{mac}}}$) and the rotational line-of-sight broadening (${v{sini}_*}$). \

\subsubsection{\textit{Gaia DR3}}

For all four objects, we obtained additional photometric and astrometric data from \textit{Gaia DR3} \citep{GAIA2016, GaiaDR3}.\

Our host stars show no evidence of being unresolved binaries, with all four targets having an astrometric excess noise of approximately zero as well as a low astrometric goodness of fit in the along-scan direction (\textit{GOF\textunderscore AL}) in \textit{Gaia DR3} \citep{Evans2018} and RUWE (re-normalised unit weight error) of approximately one. They also pass the photometric filtering processes outlined by \citet{Arenou2018} to identify blended stars.

The Gaia positions, proper motions, and absolute radial velocities of each star were used to determine the Galactic velocity components ($U_{\rm LSR},V_{\rm LSR},W_{\rm LSR}$), assuming a Local Standard of Rest of UVW=(11.1, 12.14, 7.25 $\rm km\,s^{-1}$ - \citealt{Schonrich2010}). By using the selection criteria for kinematically thin-disk ($v_{\rm tot}$ < 50 $\rm km\, s^{-1}$) and kinematically thick-disk (70 < $v_{\rm tot}$ < 180 $\rm km\, s^{-1}$) objects \citep{GAIADR2}, we conclude that three of the stars belong to the thin disk population while NGTS-23 is in-between the two populations. This appears consistent with our determination that NGTS-23 is the only metal-poor star out of our systems (see Table \ref{tbl:Spectral Properties}).\ 

\begin{table*}
\centering
\caption{The galactic velocity components and total space velocities for the four systems presented in this paper.}
\begin{tabular}{c  c  c }
\hline
Target & $\mathrm{U_{LSR}}, \mathrm{V_{LSR}},\mathrm{W_{LSR}} ({km\,s^{-1}})$ & ${v_{\rm{tot}}} ({km\,s^{-1}}) $\\
\hline\hline
NGTS-23 & -54.39, 19.95, -1.52 & 57.95 \\
NGTS-24 & 5.11, 25.74, 2.54 & 26.37 \\
NGTS-25 & -14.40, 27.52, 11.98 & 33.29 \\
HATS-54 (NGTS-22) & 31.40, -38.39, -4.24 & 49.78 \\
\hline
\end{tabular}
\label{tbl: galactic velocities}

\end{table*}

\subsubsection{SED fitting}

{We performed fits to the SEDs of each object to obtain the stellar parameters for each system. For this, we employed the Python tool \textsc{ARIADNE} \citep{Vines2022}. \textsc{ARIADNE} fits catalogue photometry to different atmospheric model grids such as Phoenix V2 \citep{Husser2013}, BT-Settl, BT-Cond, BT-NextGen \citep{Allard2012, Hauschildt1999}, \cite{Castelli2004}, and \cite{Kurucz1993}.}\

{The SEDs were modelled by interpolating the model grids in ${T_{\rm {eff}}}$ -- ${\log\,g}$ -- ${[Fe/H]}$ space. Any underestimated uncertainties in the photometry were accounted for by including an excess noise term. Gaussian priors for ${T_{\rm{eff}}}$, ${\log\,g}$, and ${[Fe/H]}$ were taken from our stellar analysis with \textsc{iSpec} (see Table \ref{tbl:Spectral Properties}), whilst Gaussian priors for the stellar radius and distance were obtained from the \textit{Gaia DR2/DR3} values, although we used an inflated value for the radius error. For NGTS-24b, we used values from \textit{Gaia} but with inflated errors as were were not able to obtain results using \textsc{iSpec}. Inflated errors were chosen so as not to unnecessarily bias the results from \textsc{ARIADNE}. In addition, $A_V$ was limited to the maximum line-of-sight value taken from galactic dust maps \citep{Schlegel1998, Schlafly2011}. The priors for the excess noise parameters were set to a normal distribution with a mean of zero and a variance equal to five times the associated uncertainty. The mass value was estimated from MIST isochrones \citep{Choi2016} using the previously derived values. The derived stellar parameters are listed in Table \ref{tbl:Spectral Properties}}.

\begin{table*}
 \caption{Stellar properties of the four systems. The second last column contains the stellar parameters for HATS-54 from the discovery paper \citep{Espinoza2019}. The final column gives the data source, for which the following references are given: 2MASS \citep{Skrutskie2006}; APASS \citep{Henden2014}; Gaia \citep{GAIA2016, GAIADR2, GaiaDR3}; WISE \citep{Wright2010}; TIC8 \citep{Stassun2019}. $^*$ denotes that, for NGTS-24b, \textsc{ARIADNE} was used instead of \textsc{iSpec} for reasons outlined in Section \ref{sec:spec analysis}.}
\begin{tabular}{G N N N N G G}
 \hline
 Property & NGTS-23 & NGTS-24 & NGTS-25 & \multicolumn{2}{c}{HATS-54 (NGTS-22)}  & {Source}\\
 & & & & Updated & {Discovery} & \\ [0.5ex] \hline\hline
 \\
 2MASS ID & J04414374-4002404 & J11141533-3754365 & J20294038-3901565 & J13223237-4441196 & -- & 2MASS \\ 
 GAIA Source ID & 4816462091938129408 & 5397086969654628864 & 6682513094425008640 & 6087996849371141248 & -- & \textit{Gaia DR2} \\
 TIC ID & 77287067 & 151920165 & 290742484 & 363851359 & -- & TIC8 \\
 R.A. & $04^h41^m43.6^s$ & $11^h14^m15.3^s$ & $20^h29^m40.3^s$ & $13^h22^m32.4^s$ & --  & \textit{Gaia DR3}\\
 DEC & $-40^{\circ}02'41.0"$ & $-37^{\circ}54'36.5"$ & $-39^{\circ}01'55.5"$ & $-44^{\circ}41'20.0"$ & --  &\textit{Gaia DR3} \\
 $\mu_{RA} ({mas\, y^{-1}})$ & 0.86±0.01 & -0.69±0.01 & 5.17±0.02 & -3.50±0.02 & -- & \textit{Gaia DR3}\\
 $\mu_{DEC} ({mas\, y^{-1})}$ & 12.27±0.02 & 0.57±0.01 & 6.03±0.02 & -7.96±0.02 & -- & \textit{Gaia DR3}\\
 Parallax (mas) & 0.95±0.01 & 1.39±0.01 & 1.90±0.02 & 1.36±0.02 & -- & \textit{Gaia DR3}\\
 \\
 \hline
 \\
 V (mag) & 14.010±0.062 & 13.192±0.063 & 14.266±0.048 & 13.914±0.027 & --  & APASS \\ 
 B (mag) & 14.578±0.024 & 13.948±0.053 & 15.146±0.052 & 14.710±0.070 & --  & APASS \\
 g (mag) & 14.291±0.028 & 13.561±0.014 & 14.656±0.048 & 14.258±0.045 & --  & APASS \\
 r (mag) & 13.940±0.049 & 13.007±0.011 & 14.000±0.049 & 13.694±0.041 & -- & APASS \\
 i (mag) & 13.845±0.043 & 12.799±0.051 & 13.844±0.092 & 13.497±0.074 & -- & APASS \\
 G (mag) & 13.9406±0.0002 & 13.0410±0.0002 & 14.0742±0.0003 & 13.7621±0.0003 & -- & \textit{Gaia DR3}\\
 J (mag) & 13.056±0.023 & 11.879±0.019 & 12.818±0.019 & 12.611±0.021 & -- & 2MASS \\
 H (mag) & 12.797±0.024 & 11.577±0.022 & 12.417±0.020 & 12.273±0.024 & -- & 2MASS \\
 K (mag) & 12.730±0.035 & 11.510±0.019 & 12.324±0.022 & 12.170±0.017 & -- & 2MASS \\
 W1 (mag) & 12.710±0.024 & 11.455±0.023 & 12.287±0.023 & 12.157±0.023 & -- & WISE \\
 W2 (mag) & 12.744±0.023 & 11.505±0.020 & 12.328±0.024 & 12.165±0.023 & -- & WISE \\
 T (mag) & 13.586±0.006 & 12.570±0.006 & 13.546±0.006 & 13.276±0.006 & -- & TIC8 \\
 \\
 \hline
 \\
 Spectral Type & F9V & G2IV & K0V & G6V & {--} &  \textsc{ARIADNE} \\
 
 ${T_{\rm{eff}}}$ & 6057±64 & 5820{\raisebox{0.5ex}{\tiny$^{+93}_{-8}$}} & 5321±58 & 5621±70 & {5702±26} & \textsc{iSpec}$^*$  \\
 
 ${log\,g}$ & 4.04±0.15 & 4.09{\raisebox{0.5ex}{\tiny$^{+0.05}_{-0.04}$}} & 4.48±0.14 & 4.20±0.20 & {4.240±0.023} & \textsc{iSpec}$^*$ \\
 
 $[Fe/H]$ & -0.26±0.11 & 0.27{\raisebox{0.5ex}{\tiny$^{+0.18}_{-0.03}$}} & 0.12±0.05 & 0.37±0.07 & {0.396±0.031} &\textsc{iSpec}$^*$ \\
 
 $[\alpha/Fe]$ & 0.07±0.17 & -- & -0.08±0.10 & -0.15±0.10 & {--}  &\textsc{iSpec} \\
 
 $v \sin i$ (km$\,\rm{s}^{-1}$) & 1.46±1.08 & -- & 1.96±0.68 & <2.50 & {3.83±0.42} &\textsc{iSpec} \\
 
 ${v_{\rm{mic}}}$ (km$\,\rm{s}^{-1}$) & 1.24±0.20 & -- & 0.89±0.21 & 0.88±0.23 & {0.948±0.034} & \textsc{iSpec} \\
 
 ${v_{\rm{mac}}}$ (km$\,\rm{s}^{-1}$) & 5.71±0.84 & -- & 3.24±0.78 & 4.49±0.68 & {3.61±0.11} & \textsc{iSpec} \\

 $R_* (R_{\odot})$ & 1.24±0.03 & 1.64{\raisebox{0.5ex}{\tiny$^{+0.03}_{-0.09}$}} & 0.86{\raisebox{0.5ex}{\tiny$^{+0.02}_{-0.05}$}} & 1.23±0.02 & {1.317±0.036} & \textsc{ARIADNE} \\
 
 $M_* (M_{\odot})$ & 1.01{\raisebox{0.5ex}{\tiny$^{+0.06}_{-0.07}$}} & 1.26{\raisebox{0.5ex}{\tiny$^{+0.05}_{-0.21}$}} & 0.91{\raisebox{0.5ex}{\tiny$^{+0.03}_{-0.04}$}} & 1.05±0.05 & {1.097±0.022}  & \textsc{ARIADNE}\\
 
 $\rho_* (\rm{g\,cm^{-3}})$ & 0.755{\raisebox{0.5ex}{\tiny$^{+0.027}_{-0.034}$}} & $0.406\pm0.059$ & 2.219{\raisebox{0.5ex}{\tiny$^{+0.085}_{-0.110}$}}  & $0.795\pm0.052$ & {0.678±0.054} & Joint Model \\
 
 $L_* (L_{\odot})$ & 1.920{\raisebox{0.5ex}{\tiny$^{+0.093}_{-0.106}$}} & 2.792{\raisebox{0.5ex}{\tiny$^{+0.215}_{-0.289}$}} & 0.511{\raisebox{0.5ex}{\tiny$^{+0.045}_{-0.052}$}} & 1.302{\raisebox{0.5ex}{\tiny$^{+0.049}_{-0.055}$}} & {1.631{\raisebox{0.5ex}{\tiny$^{+0.114}_{-0.076}$}}} & \textsc{ARIADNE} \\
 
 $d$ (pc) & 991{\raisebox{0.5ex}{\tiny$^{+19}_{-23}$}} & 725{\raisebox{0.5ex}{\tiny$^{+33}_{-28}$}} & 517{\raisebox{0.5ex}{\tiny$^{+18}_{-22}$}} & 720{\raisebox{0.5ex}{\tiny$^{+11}_{-12}$}} & {769±21} & \textsc{ARIADNE} \\
 
 $A_{v}$\ (mag) & 0.019{\raisebox{0.5ex}{\tiny$^{+0.029}_{-0.016}$}} & 0.249{\raisebox{0.5ex}{\tiny$^{+0.035}_{-0.067}$}} & 0.077{\raisebox{0.5ex}{\tiny$^{+0.061}_{-0.054}$}} & 0.114{\raisebox{0.5ex}{\tiny$^{+0.042}_{-0.038}$}} & {0.279±0.018} & \textsc{ARIADNE}\\
 
  Age (Gyrs) & 6.39{\raisebox{0.5ex}{\tiny$^{+1.93}_{-1.60}$}} & 4.77{\raisebox{0.5ex}{\tiny$^{+3.56}_{-0.91}$}} & 3.49{\raisebox{0.5ex}{\tiny$^{+3.49}_{-3.37}$}} & 7.96{\raisebox{0.5ex}{\tiny$^{+2.97}_{-2.26}$}} & {6.60±0.76} & \textsc{ARIADNE} \\
  
 \hline
 \end{tabular}

 \centering
 \label{tbl:Spectral Properties}
\end{table*}

\subsubsection{Stellar Rotation}

As well as modelling the stellar parameters, we also wished to constrain the stellar rotation period. Coupled to a knowledge of the stellar radius and ${v\rm{sini}_*}$, determination of the stellar rotation period can then allow constraints to be placed on the inclination angle of the stellar rotation axis. This can enable misaligned star-planet systems to be identified \citep{Watson2010}. \

The first method used to constrain the rotation period was to spectroscopically measure the chromospheric activity of the star. Using the formulae described in \citet{Lovis2011}, the log$\,\rm{R'_{HK}}$ was measured from the co-added HARPS spectra used in Section \ref{sec:spec analysis}. Therefore, the only system we were not able to obtain a log$\,\rm{R'_{HK}}$ value for was NGTS-24b.  From these co-added spectra, we found log$\,\rm{R'_{HK}}$ values of -5.43, -4.61 and -4.74 for NGTS-23, NGTS-25 and HATS-54, respectively. Using the relation from \citet{Noyes1984}, the most probable stellar rotation periods are predicted to be 27\,days for NGTS-23, 24\,days for NGTS-25 and 29\,days for HATS-54.\

Our second approach used the ${v\,{\sin i}_*}$ of the star measured from \textsc{iSpec}. By assuming that the stellar inclination angle, $i_*$, is 90\textdegree\ (i.e. $\sin i_* $=1), an upper limit can be placed on the stellar rotation period when the stellar radius is known. NGTS-23 and NGTS-25 were found to have a ${v\,{\sin i}_*}$ of 1.46±1.08 km$\,\rm{s}^{-1}$ and 1.96±0.68 km$\,\rm{s}^{-1}$ respectively, which correspond to upper limits of the rotation periods of 43±32\,days and 22±8\,days. We were only able to obtain an upper-limit to the ${v\,{\sin i}_*}$ for HATS-54, finding it to be <2.50 km$\,\rm{s}^{-1}$. This corresponds to an upper limit of 25\,days. However, if the value of ${v\,{\sin i}_*}$ is lower then this upper limit will increase, which means that HATS-54 is likely a slow rotator.  These are consistent with the previous estimates from the log$\,\rm{R'_{HK}}$ and appear consistent with the age and spectral types of the stars. \

Finally, we searched the Generalised Lomb-Scargle periodograms \citep{Lomb1976, Scargle1982, Zechmeister2009} of the NGTS photometry after the transits were removed (see appendix \ref{fig:NGTS periodograms} for plots). NGTS-23 has significant peaks at approximately 15\,days and 33\,days. The 33\,day peak is more consistent with the previous estimate of 27\,days from the log$\,\rm{R'_{HK}}$. The most significant peak for NGTS-24 is at approximately 30\,days. However, it is possible that this peak is indicative of the moon contaminating the photometry. However, given that the next most significant peak is about 47\,days, it appears likely that NGTS-24 is slowly rotating. NGTS-25 has peaks in the periodogram at approximately 14\,days and 28\,days, which could be caused by moonlight in the photometry. There is also a significant peak between 40 and 50\,days but this is not consistent with the upper limit from the ${v\,\sin i_*}$ and the estimate of 24\,days from the log$\,\rm{R'_{HK}}$ so it is unlikely to correspond with the rotation period. Finally, HATS-54 has numerous significant peaks. There is a significant peak at roughly 27\,days, which is broadly consistent with the previous estimates. We conclude, like the other stars presented in this paper, that HATS-54 is slowly rotating. \

As a final check, we explored alternative methods, including autocorrelation and GP regression in addition to LS periodogram \citep{Gillen20}, via the Rotation Toolkit (RoTo)\footnote{\href{https://github.com/joshbriegal/roto}{https://github.com/joshbriegal/roto}}, but do not detect clear rotation patterns beyond the aforementioned suggestive signals.

\subsubsection{Stellar Properties}

\begin{figure}
    \centering
    \includegraphics[width=0.475\textwidth]{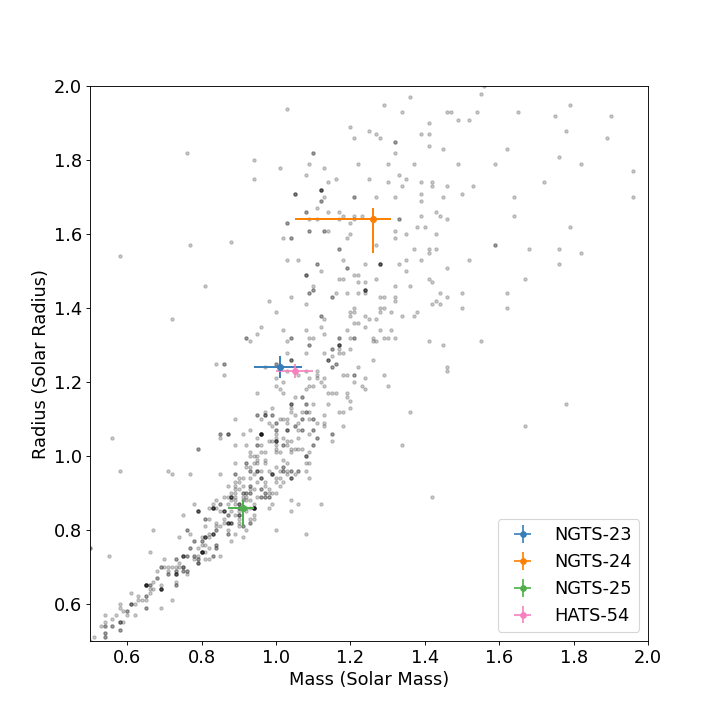}
    \caption{{Stars that host exoplanets with defined radii and masses (grey points) with errorbars suppressed for clarity. The stars presented in this paper are plotted as coloured points with errorbars. It can be seen that three of the stars presented in this paper lie in relatively sparse regions of (stellar) mass-radius space while NGTS-25 lies in the general population. The stellar sample was obtained from the NASA Exoplanet Archive (\url{https://exoplanetarchive.ipac.caltech.edu/}).} }
    \label{fig: starswithplanets}
\end{figure}

\begin{figure}
    \centering
    \includegraphics[width=0.475\textwidth]{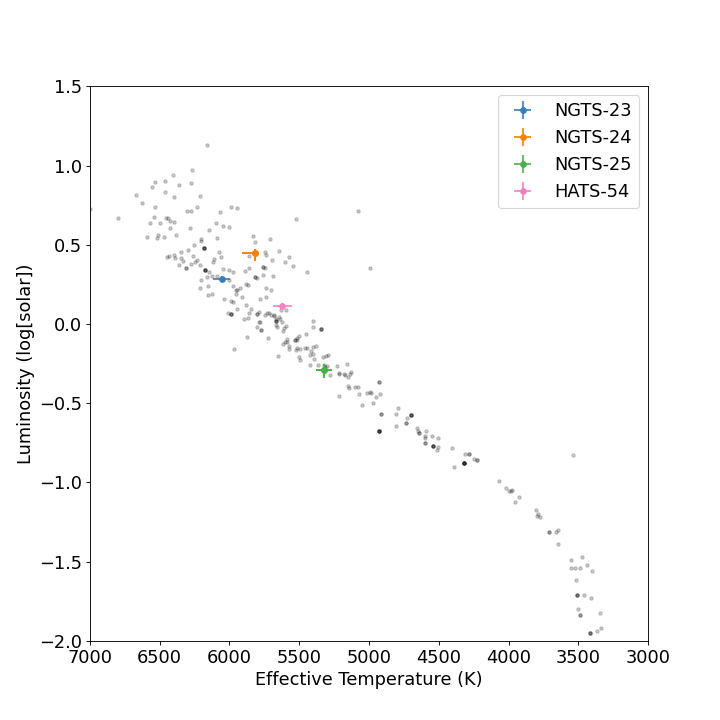}
    \caption{A Hertzsprung-Russell diagram for the same stellar sample described in Figure \ref{fig: starswithplanets}. The errorbars are suppressed for clarity. The stars presented in this paper are plotted as coloured points with errorbars. Of particular note is NGTS-24, which appears to be beginning to transition off the main-sequence. HATS-54 also appears to be towards the end of its main-sequence lifetime. } 
    \label{fig: HRdiagram}
\end{figure}

Using the parameters obtained for the stellar radii, masses, effective temperatures and luminosities, we can compare our host stars to the general population of stars that host planets, where the planets have defined masses and radii. This is a sample of approximately 950 stars, and a stellar mass-radius diagram and a Hertzsprung-Russell diagram can be seen in Figures \ref{fig: starswithplanets} and \ref{fig: HRdiagram}, respectively.\

{From these figures, it appears that NGTS-25 is on the main-sequence and lies in the general population of stars. NGTS-23 and HATS-54 both occupy a sparse region of mass-radius parameter space. Finally, NGTS-24 is in an even less populated region of mass-radius space and appears to be transitioning off the main-sequence if it has not already done so. This appears a reasonable conclusion given its age and spectral type (see Table \ref{tbl:Spectral Properties}). Therefore, it is an interesting target given the rarity of hot Jupiters around sub-giant stars \citep{Villaver2009, Hansen2010, Schlaufman2013}. We also conclude that HATS-54 and NGTS-23 are both towards the end of their lifetimes on the main-sequence and will evolve off the main-sequence within a few Gyrs. Again, this seems to be in good agreement with their ages and spectral type. These conclusions are further supported from the findings in Section \ref{sec:decay}, in which all three stars show significant evolution within relatively short periods of time.}

\subsection{Global Modelling}\label{sec:global}

We determined the physical parameters of the systems, including planet radius and mass, via a simultaneous fit to the photometric and spectroscopic data using the open-source astronomy software package \textsc{allesfitter} \citep{allesfitter-code, allesfitter-paper}. \textsc{allesfitter} combines the packages \textsc{ellc} (light curve and RV models; \citealt{Maxted2016}), \textsc{emcee} (MCMC sampling; \citealt{Foreman-Mackey2013}), \textsc{dynesty} (nested sampling; \citealt{Speagle2020}), and \textsc{celerite} (Gaussian Process (GP) models; \citealt{Foreman-Mackey2017}). This combination allows \textsc{allesfitter} to model a variety of signals, including stellar variability and transit-timing variations.\

Given the length of time for the initial MCMC fits to converge, a nested sampling approach was taken to produce the global fits. However, quick MCMC fits (100 walkers and 5000 steps) to the NGTS data were used to refine priors on parameters such as transit epoch and transit depth. Gaussian priors were placed on the stellar mass, radius and temperature using the results from the SED fitting (see Table \ref{tbl:Spectral Properties}). We adopted a quadratic limb darkening law for the photometric data as parameterized in \citet{Kipping2013}. Gaussian processes with a Matern 3/2 kernel were used to fit any out-of-transit variability and for this reason we removed lightcurves from the final models that did not have a sufficient out-of-transit baseline. Therefore, for NGTS-23b we did not include the EulerCam data or the SAAO 1.0\,m data in our final fits, while for HATS-54b we did not use the EulerCam data taken on the {2019/03/25} or either night of the photometry taken by the SAAO 1.0\,m in our final fits. Transits that were not included in the final models can be found in appendix \ref{fig:appendix phot plots} for completeness. However, for HATS-54b we made use of both photometry and radial velocity measurements from \citet{Espinoza2019} in our global models. Furthermore, as multiple spectrographs were used to obtain radial velocity measurements for each system, we fit for a constant instrumental velocity offset for each spectrograph. \

For each system, we ran multiple fits before performing our final fit. We first tested removing  RV points that were measured during transit as these points could have been affected by the Rossiter-McLaughlin effect \citep{Rossiter1924, Mclaughlin1924}. We found that removing these points did not significantly alter the results and therefore these in-transit RV points were included in the final fits.\

Finally, we ran two additional fits for each planet -- one where the planet was fixed on a circular orbit (i.e. $e = 0$), and another where the eccentricity was allowed to vary freely. Each of the latter fits resulted in non-zero eccentricity values. However, many small eccentricity values are spurious and \citet{Lucy1971} define a probabilistic test to determine whether a small e is statistically significant from 0. By adopting a 5$\%$ level of significance, they find that if the condition $e > 2.45\,\sigma_{e}$ is satisfied (where $\sigma_{e}$ is the observational uncertainty on $e$), then one can be confident that the measured eccentricity is real. We find that all of our measurements fail to meet this criteria, and we therefore fixed the eccentricity to be zero in our final fits. This is in-line with both theoretical predictions, and the observed population of hot Jupiters, where orbital eccentricity is often consistent with zero \citep[e.g.][]{Anderson2012}. \

We use \textsc{allesfitter} to fit our data for the key physical parameters of each system, including the planet’s orbital period $P$, the radius ratio $R_P$/$R_*$, and the radial velocity semi-amplitude $K$. A full list of physical properties and their values for each planet can be found in Table \ref{tbl: planet params}. Figure \ref{fig:photmetry plots} shows the phase-folded lightcurves for each object. Similarly, the phase-folded radial velocity data for each star is shown in Figure \ref{fig:rv plots}.

\begin{table*}
    \caption{{Planetary parameters for the four systems presented in this paper. The final 2 columns display the updated planetary parameters for HATS-54b/NGTS-22b alongside those from the discovery paper \citep{Espinoza2019}.}}
    \centering
    \begin{tabular}{G F K F K K }
    \hline
        Property & NGTS-23b & NGTS-24b  & NGTS-25b & \multicolumn{2}{c}{HATS-54b / NGTS-22b} \\
        & & & & Updated & {Discovery} \\
        \hline\hline
        \\
            $P$ (days) & $4.0764326\pm0.0000041$ & $3.4678796\pm0.0000070$ & $2.8230930\pm0.0000033$ & {2.5441765±0.0000024} & {2.5441828±0.0000043} \\ 
            
            $T_C$ (BJD) & $2458431.77134\pm0.00048$ & $2458100.6229\pm0.0011$ & $2458204.26373\pm0.00050$ & {2457780.01023±0.00061} & {2457780.01102±0.00089}  \\
            
            $T_{14}$ (hours) & $3.919\pm0.031$ & 2.929{\raisebox{0.5ex}{\tiny$^{+0.095}_{-0.083}$}} & $2.466\pm0.025$ & {2.468{\raisebox{0.5ex}{\tiny$^{+0.044}_{-0.041}$}}} & {2.5008±0.0456} \\
            
            $R_{\rm P}/R_*$ & 0.1048{\raisebox{0.5ex}{\tiny$^{+0.0011}_{-0.0011}$}} & 0.0768{\raisebox{0.5ex}{\tiny$^{+0.0028}_{-0.0023}$}} & 0.1231{\raisebox{0.5ex}{\tiny$^{+0.0011}_{-0.0013}$}} & {0.0848±0.0014} & {0.0832±0.0025} \\
            
            $a/R_*$ & 8.72{\raisebox{0.5ex}{\tiny$^{+0.10}_{-0.13}$}} & 6.37{\raisebox{0.5ex}{\tiny$^{+0.29}_{-0.31}$}} & 9.78{\raisebox{0.5ex}{\tiny$^{+0.12}_{-0.16}$}} & {6.48±0.14} & {6.15±0.16} \\
            
            $b$ & 0.134{\raisebox{0.5ex}{\tiny$^{+0.094}_{-0.091}$}} & 0.821±0.026 & 0.113{\raisebox{0.5ex}{\tiny$^{+0.097}_{-0.072}$}} & {0.715{\raisebox{0.5ex}{\tiny$^{+0.024}_{-0.026}$}}} & {0.740{\raisebox{0.5ex}{\tiny$^{+0.022}_{-0.023}$}}} \\
            
            $i$ $(\degree)$ & 89.12{\raisebox{0.5ex}{\tiny$^{+0.53}_{-0.64}$}} & 82.61{\raisebox{0.5ex}{\tiny$^{+0.52}_{-0.61}$}} & 89.34{\raisebox{0.5ex}{\tiny$^{+0.42}_{-0.58}$}} & {83.67±0.34} & {83.08±0.36} \\
            
            $e$ & 0 (fixed) & 0 (fixed) & 0 (fixed) & 0 (fixed) & {<0.126} \\
            
            $K$ $(\rm{km\,s^{-1}})$ & $0.078\pm0.011$ & 0.062{\raisebox{0.5ex}{\tiny$^{+0.011}_{-0.010}$}} & $0.100\pm0.006$ & {0.108±0.006} & {0.105±0.014} \\
            
            $R_{P} ({R_{\rm J}})$ & $1.267\pm0.030$ & 1.214{\raisebox{0.5ex}{\tiny$^{+0.058}_{-0.062}$}} & 1.023{\raisebox{0.5ex}{\tiny$^{+0.035}_{-0.052}$}} & {1.015±0.024} & {1.067±0.052}  \\
            
            $M_P ({M_{\rm J}})$ & 0.613±0.097 & 0.520{\raisebox{0.5ex}{\tiny$^{+0.120}_{-0.110}$}} & 0.639{\raisebox{0.5ex}{\tiny$^{+0.058}_{-0.052}$}} & {0.753±0.057} & {0.76±0.10} \\
            
            $\rho_P$ $(\rm{g\,cm^{-3}})$ & 0.373{\raisebox{0.5ex}{\tiny$^{+0.070}_{-0.064}$}} & 0.359{\raisebox{0.5ex}{\tiny$^{+0.120}_{-0.087}$}} & 0.892{\raisebox{0.5ex}{\tiny$^{+0.099}_{-0.089}$}} & {0.940{\raisebox{0.5ex}{\tiny$^{+0.120}_{-0.110}$}}} & {0.77±0.16} \\
            
            $a$ (au) & 0.0504±0.012 & 0.0479{\raisebox{0.5ex}{\tiny$^{+0.0028}_{-0.0030}$}} & 0.0388{\raisebox{0.5ex}{\tiny$^{+0.0014}_{-0.0020}$}} & {0.0370±0.0010} & {0.03763±0.00024} \\
            
            ${T_{\rm {eq}}}$ (K) & 1327{\raisebox{0.5ex}{\tiny$^{+17}_{-16}$}} & 1499{\raisebox{0.5ex}{\tiny$^{+41}_{-36}$}} & 1101{\raisebox{0.5ex}{\tiny$^{+15}_{-14}$}} & {1429±24} & {1625±22}\\
            
            $\rm{Irradiation}\,(\rm{W\,m^{-2}})$ & (1.03±0.07)$\times10^6$ & (1.66{\raisebox{0.5ex}{\tiny$^{+0.24}_{-0.26}$}})$\times10^6$ & (4.62{\raisebox{0.5ex}{\tiny$^{+0.63}_{-0.57}$}})$\times10^5$ & {(1.29{\raisebox{0.5ex}{\tiny$^{+0.08}_{-0.09}$}})$\times10^6$}  & {(1.57{\raisebox{0.5ex}{\tiny$^{+0.09}_{-0.08}$}})$\times10^6$} \\
            \hline
    \end{tabular}

    \label{tbl: planet params}
\end{table*}

\section{Results and discussion}\label{Sec:Results}
\begin{figure}
    \centering
    \includegraphics[width=0.475\textwidth]{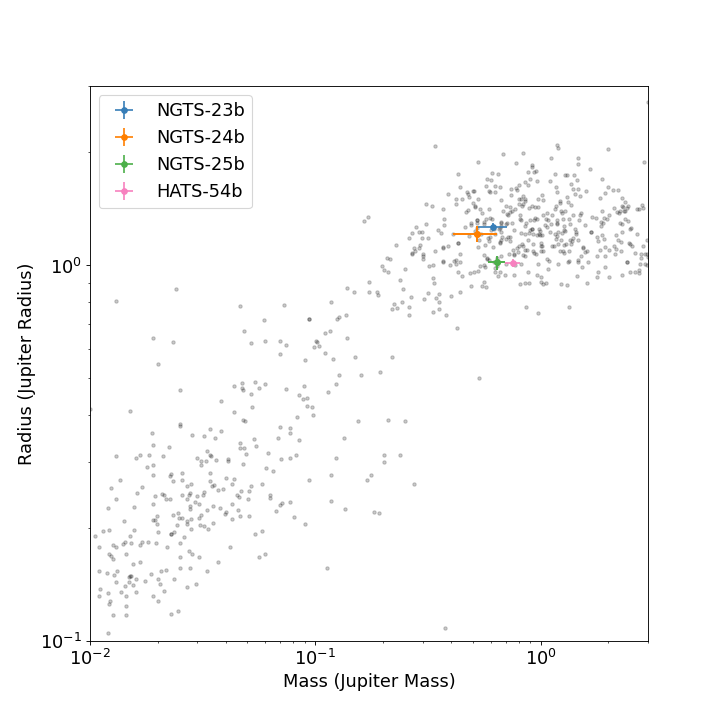}
    \caption{{Exoplanets with defined masses and radii (grey points) with errorbars suppressed for clarity. The planets presented in this paper are plotted as coloured points with errorbars. All four planets presented are found to lie within the general population of hot Jupiters. The exoplanet sample was obtained using the NASA Exoplanet Archive (\url{https://exoplanetarchive.ipac.caltech.edu/})}} 
    \label{fig:massvsradius}
\end{figure}

All the planets presented in this work have radii between 1.02 - 1.27 ${R_{\rm J}}$ and masses between 0.52 - 0.75 ${M_{\rm J}}$. This places them in the general population of hot Jupiters (Figure \ref{fig:massvsradius}), with planetary bulk densities ranging from 0.359 $\rm{g\,cm^{-3}}$ to 0.940 $\rm{g\,cm^{-3}}$. \

{Our stellar parameters for HATS-54 (NGTS-22) are generally in good agreement with those from \citet{Espinoza2019}. $T_{\mathrm{eff}}$, ${\log\,g}$, $M_*$, $[Fe/H]$ and ${v_{\rm {mic}}}$ all agree within the error bars. The radius values are within $2\sigma$ while the luminosity values are within $3\sigma$. There are significant differences between the ${v\,{\sin i}_*}$ and ${v_{\rm {mac}}}$ values reported in the discovery paper and the values determined in this work. However, as \citet{Espinoza2019} assumed a ${v_{\rm {mac}}}$ value that is lower than the value in this paper, this will result in a larger ${v\,{\sin i}_*}$ due to the degeneracy between the two properties. Given the age and spectral type of HATS-54, our slowly rotating value seems more appropriate. There was also disagreement on the extinction values but the value from \citet{Espinoza2019} is high given the limit from both \citet{Schlegel1998} and \citet{Schlafly2011}.}

{The planetary parameters obtained for HATS-54b (NGTS-22b) are in good agreement with those reported in the discovery paper. Most of the values agree within error bars but $a/R_*$ and the impact parameter, $b$, differ by less than $2\sigma$. However, the main difference is that \citet{Espinoza2019} could only constrain $e<0.126$ whereas we fixed $e=0$ as non-zero eccentricity is likely spurious given that our eccentric models returned an eccentricity value of ${e}=0.023{\raisebox{0.5ex}{\tiny$^{+0.033}_{-0.017}$}}$.} \

\subsection{Orbital Decay Time-scales}\label{sec:decay}

Given the late age of two of our systems (NGTS-24 and HATS-54), we thought it useful to calculate the orbital decay time-scales for our planets. We use the following equation to calculate the rate of orbital decay \citep{Goldreich1966, Ogilvie2014, Patra2020}:

\begin{equation}
    \frac{dP}{dt}=-\frac{27\pi}{2Q'_*}\left(\frac{M_p}{M_*}\right)\left(\frac{R_*}{a}\right)^5
\end{equation}

\noindent where $Q'_*=3Q_*/2k_2$. $Q_*$ and $Q'_*$ are the tidal quality factor and the reduced tidal quality factor, respectively, while $k_2$ is the dimensionless quadrupolar Love number. We also only consider the equilibrium tide within the convection zone in this analysis as none of our systems satisfy the criterion for the dynamical tide to be considered. This is because the inertial waves (that constitute the dynamical tide) are only excited if the orbital period is twice the stellar rotation period (i.e. ${P_{\rm{orb}}}>P_*/2$). Therefore, for our analysis, we can use $Q'_*$ as defined in \citet{Strugarek2017}.\footnote{This analysis requires the depth of the convection zone. Therefore the radius and mass of the core were determined using \url{http://www.astro.wisc.edu/~townsend/static.php?ref=ez-web}}\

Using the orbital period and the rate of decay, we are able to make a prediction for the length of time for the planetary orbit to decay completely (i.e. where its orbital period is zero). This assumes a constant rate of decay and thus will likely over-estimate the time taken for the orbit to decay. We also have ignored the effects of stellar wind which would likely decrease the inspiral time as well. Finally, as the star evolves off the main-sequence the tidal quality factor can change by orders of magnitude, which  dramatically impacts the decay time. This could in-part explain the lack of hot Jupiters observed around sub-giant stars \citep{Villaver2009, Hansen2010, Schlaufman2013}.\

A summary of all the estimated decay time-scales can be found in Table \ref{tbl:decay times}. Although we have estimated the rate of decay for these systems, we are likely unable to measure these due to their long decay time-scales. For example, the observed orbital decay rate of WASP-12b is several orders of magnitude higher than for any of the stars presented here \citep{Maciejewski2016, Patra2017, Yee2020}. The long orbital decay time-scales, while an estimation, suggest that all four planets are in a relatively stable configuration while the respective host star remains on the main-sequence.\

However, we see that when the star evolves, the orbital decay time-scale rapidly decreases, which agrees with the observations that there is a lack of hot Jupiters around sub-giant stars. Of particular interest is NGTS-24b as its host star will soon evolve off the main-sequence if it is not already doing so. There is some tentative evidence to suggest that NGTS-24b is already beginning to evolve off the main-sequence (see Table \ref{tbl:Spectral Properties} and Figure \ref{fig: HRdiagram}). Therefore, we expect that the star will change significantly within 500 million years causing the inspiral time to reduce significantly. This leads to the planet being engulfed within 600 million years. For comparison, HATS-54 will take 2 billion years before it evolves to a stage where it will have a similar inspiral time.

\begin{table*}
    \caption{Summary of the decay time-scales for the four planets presented in this paper. In column 2, "Age", we list two values for each star. The first is the stellar age estimation from Table \ref{tbl:Spectral Properties}.  For comparison, the second value indicates the age at which  the star is estimated to begin evolving off of the main-sequence. For NGTS-25b, this later age was decided to be 10 Gyrs older than its determined age. ${M_{\rm{c}}}$ and ${R_{\rm c}}$ are the mass and radius of the core respectively. $Q'_*$ is the reduced tidal quality factor, $\frac{dP}{dt}$ is the rate of orbital decay and $t_a$ is the associated time for the planetary orbit to decay. The stellar rotation period was assumed to be 30 days in each scenario. We note that the assumption of an incorrect stellar rotation period does not affect the results much unless the true period is $<<$1day.}
    \centering
    \begin{tabular}{c c c c c c c c c c c }
    \hline
        Star & Age (Gyrs) & $M_*$ (${M_\odot}$) & $R_*$ (${R_\odot}$) & $L_*$ (${L_\odot}$) & ${M_{\rm c}}$ (${M_\odot}$) & ${R_{\rm c}}$ (${R_\odot}$) & $Q'_*$ & $\frac{dP}{dt}$ & $t_a$ (Gyrs) \\
        \hline\hline
        
        \multirow{2}{*}{NGTS-23} & 6.39 & 1.01 & 1.24 & 1.920 & 1.00 & 0.95 &  6.49$\times10^6$ &  -7.43$\times10^{-14}$ & 150 \\

        & 8.89 & 1.01 & 2.32 & 3.371 & 0.63 & 0.94 & 3.15$\times10^5$ & -3.51$\times10^{-11}$ & 0.32\\
        \hline
        \multirow{2}{*}{NGTS-24} & 4.77 & 1.26 & 1.64 & 2.792 & 1.23 & 1.18 & 2.43$\times10^6$ & -7.02$\times10^{-13}$ & 14 \\
        
        & 5.27 & 1.26 & 2.94 & 4.485 & 0.52 & 0.93 & 3.16$\times10^5$ & -1.00$\times10^{-10}$ & 0.09        \\

        \hline
        \multirow{2}{*}{NGTS-25} & 3.49 & 0.91 & 0.86 & 0.511 & 0.88 & 0.57 &  1.06$\times10^7$ & -3.12$\times10^{-14}$  & 248 \\
        & 13.49 & 0.91 & 1.06 & 0.912 & 0.87 & 0.70 & 4.60$\times10^6$ & -2.05$\times10^{-13}$ & 38\\
        \hline        
        HATS-54 & 7.96 & 1.05 & 1.23 & 1.302 & 1.02 & 0.89 & 4.17$\times10^6$ & -6.05$\times10^{-13}$ & 12        \\

        (NGTS-22)& 9.96 & 1.05 & 2.11 & 2.360 & 0.66 & 0.88 & 4.98$\times10^5$ & -7.54$\times10^{-11}$ & 0.09\\
        
        \hline

    \end{tabular}

    \label{tbl:decay times}
\end{table*}

\subsection{Inflation}\label{sec:inflation}

{Many hot Jupiters that are highly irradiated are observed to have low densities \citep{Laughlin2011, Weiss2013, Thorngren2018}. Previous studies have suggested that this is driven by inflation mechanisms that become effective above an incident flux of approximately $2\times10^5\, \rm{W\,m^{-2}}$ \citep{Demory2011, Miller2011}.}\

{We calculated the irradiation received by each planet and found that all four are irradiated above this threshold (see Table \ref{tbl: planet params}). Therefore, we would expect that our planets might exhibit larger than expected radii due to inflation.}\

{We follow the procedure laid out by \citet{Tilbrook2020} and \citet{Costes2020}, who utilise the work of \citet{Baraffe2008} (hereafter B08) and \citet{Sestovic2018} (hereafter S18) to determine whether a planet is inflated. Using the work from B08 we obtain estimated radii for non-inflated planets that we can compare with expected inflated radii from S18.} \

{B08 produced planetary evolution models that account for previous limitations, such as quantifying the effect of metallicity. Notably, radius decreases with increasing heavy element mass so by assuming a low metallicity of 0.02 - 0.1 for each planet, we are able to provide an upper limit for the non-inflated radii. Differences between the predicted value and the observed radius could indicate that inflation mechanisms have affected the radius.} \

S18 implement hierarchical Bayesian modelling and a forward model to infer relationships between incident flux and radius. The increase in radius, $\Delta R$, is also dependent on the mass of the planet. All four of the planets presented in this paper fall into the same range (0.37 - 0.98 ${M_{\rm J}}$), and we therefore use the corresponding equation to derive the expected radius increase due to inflation for each:

\begin{equation}
\centering
    \Delta R =0.70 \times (\mathrm{log_{10}F} - 5.5).
\end{equation}

\noindent The baseline radius in S18 is ${R_{\rm {base}}}=(0.98\pm 0.04)\: {R_{\rm J}}$ for this mass regime. Therefore, the inflated radius is calculated as:

\begin{equation}
    {R_{\rm{inflated}}}=0.98 + \Delta R .
\end{equation}

\noindent These inflated radii can then be compared with the derived radii from our global models.\

For NGTS-23b, the observed radius is consistent with the inflated radius as predicted by S18 within error-bars. It is also larger than the estimated non-inflated radius by more than $6\sigma$. It is therefore reasonable to conclude that NGTS-23b is inflated. NGTS-24b has a radius greater than the estimated non-inflated radius, but it is not as large as the expected inflated radius. Although there is less than a $3\sigma$ difference between the observed radius and the expected non-inflated radius, the model from S18 fails to predict the behaviour. However, due to the high irradiation and the low planetary density it is likely that this planet is inflated. Therefore, our results could be indicative that density is a more suitable parameter than planetary radius to determine whether a planet is inflated. HATS-54b has a radius that is consistent with the non-inflated model and disagrees with the inflated prediction by more than $6\sigma$. Therefore, it appears that HATS-54b is not inflated. Finally, NGTS-25b is consistent with both inflated and non-inflated predictions so we are not able to draw any firm conclusions for this planet. However, this is not altogether surprising given that NGTS-25b has a lower level of irradiation compared to the other planets presented in this paper. A summary of the inflation results can be found in Table \ref{tbl:inflation}.\

\begin{table*}
    \caption{Quantification of the inflation of the four planets. Note the masses in this table are not the true planetary masses but are the masses used to calculate the non-inflated radius. $\Delta$R is the expected change in radius caused by inflation mechanisms.}
    \centering
    \begin{tabular}{c c c c c c c }
    \hline
        Planet & Irradiation $(\rm{W\,m^{-2}})$ & $M_*$ (${M_{\rm J}}$) & $R_{\rm{non-inflated}}$ (${R_{\rm J}}$) & $\Delta R$ (${R_{\rm J}}$) &  $R_{\rm{inflated}}$ (${R_{\rm J}}$) & $R_{\rm {observed}}$ (${R_{\rm J}}$)\\
        \hline\hline
        
        NGTS-23b & (1.03±0.07)$\times10^6$ & 0.5 & 1.017-1.093 & 0.36{\raisebox{0.5ex}{\tiny$^{+0.01}_{-0.02}$}} & 1.34±0.04 & 1.27±0.03
        \\
        
        NGTS-24b & (1.66{\raisebox{0.5ex}{\tiny$^{+0.24}_{-0.26}$}})$\times10^6$ & 0.5 & 1.009-1.072 & 0.50±0.03 & 1.48±0.05 & 1.21±0.06
        \\
        
        NGTS-25b & (4.62{\raisebox{0.5ex}{\tiny$^{+0.63}_{-0.57}$}})$\times10^5$ & 0.5-1.0 & 1.009-1.186 & 0.12±0.03 & 1.10±0.05 & 1.02{\raisebox{0.5ex}{\tiny$^{+0.04}_{-0.05}$}}
        \\
              
        HATS-54b & (1.29{\raisebox{0.5ex}{\tiny$^{+0.08}_{-0.09}$}})$\times10^6$ & 0.5-1.0 & 1.009-1.106 & 0.42±0.01 & 1.40±0.04 & 1.02±0.02
        \\
        \hline

    \end{tabular}

    \label{tbl:inflation}
\end{table*}

\begin{figure}
\centering
\includegraphics[width=0.475\textwidth]{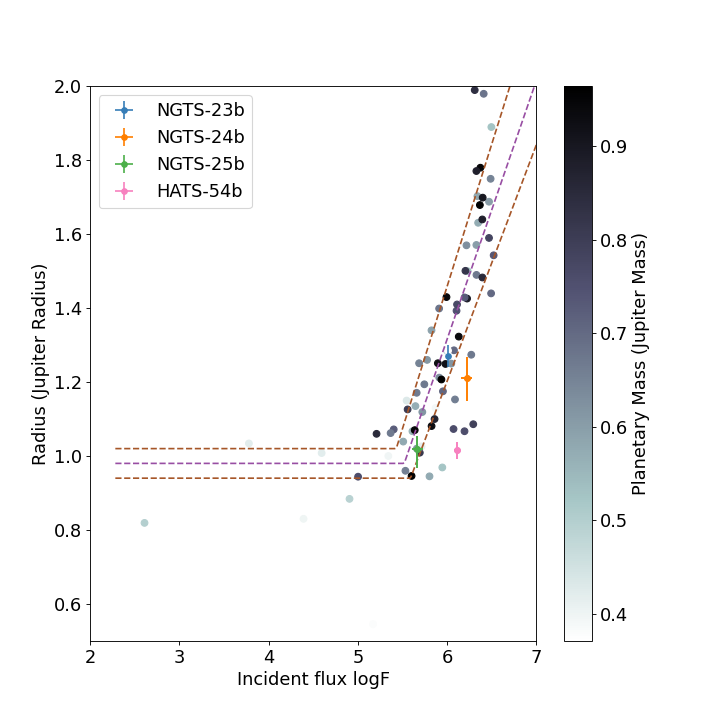}
\caption{Exoplanets from the NASA Exoplanet Archive (\url{https://exoplanetarchive.ipac.caltech.edu/}) with masses between 0.37 - 0.97 ${M_{\rm J}}$. The purple dashed line shows the predicted `inflated' radii \citep{Sestovic2018} with the associated standard deviation shown as dashed brown lines. The hot Jupiters presented in this paper are plotted as coloured points with errorbars.} 
\label{fig:sestovic}
\end{figure}

{Similar to the results from \citet{Tilbrook2020}, our results are in tension with the claim from S18 that there is not a population of non-inflated hot Jupiters in the mass range 0.37-0.98 ${M_{\rm J}}$. Although all of the incident fluxes are larger than the threshold for inflation \citep[$2\times 10^5\, \rm{W\,m^{-2}}$ -][]{Demory2011, Miller2011}, it appears that HATS-54b is not inflated. It is also unclear whether NGTS-24b and NGTS-25b are inflated. Given that this is simply the minimum limit at which planets begin to appear inflated and that the incident flux for NGTS-25b is not much greater than this limit, it is not too surprising that NGTS-25b is not inflated. However, as new evidence suggests that gaseous planets may experience ‘reinflation’ as the star evolves off the main-sequence \citep{Hartman2016, Komacek2020,Thorngren2020, Thorngren2021}, it is somewhat surprising that NGTS-24b cannot be confirmed to be inflated as it experiences relatively high levels of irradiation. Therefore, NGTS-24b may be an interesting candidate to study to better understand the impact of additional parameters on the mechanisms driving inflation.}  \

In particular, metallicity might play an important role as, for the systems reported in this paper, the only planet orbiting a metal-poor star was also the only planet that agreed with the inflated radius predictions. This agrees with the expectation that radii decrease as the fraction of heavy elements in a planet increases \citep{Thorngren2016}. It is already known that hot Jupiter radii vary with age \citep{Baraffe2008, Miller2009, Thorngren2018}, and additionally, differing system ages could indicate a difference in planetary migration times, and therefore variations in the level of stellar irradiation compared to what they currently experience. The detailed effects of additional parameters, such as metallicity, eccentricity and system age, are not yet fully understood. However, they could potentially explain why planets may exhibit different levels of radius inflation despite receiving similar levels of irradiation.\

Furthermore, the number of known hot Jupiters has increased greatly since the publication of S18, therefore it would be beneficial to have an updated model with an aim of understanding the impact of various factors as this could describe the inflation relationships more accurately. Understanding the dependence on these parameters is crucial to determining the underlying inflation mechanisms.\

\section{Conclusions}\label{Sec:Conclusions}

We report the discovery of three new hot Jupiters and the independent detection of HATS-54b. Each planet was originally identified from photometry from the Next Generation Transit Survey (NGTS), and was confirmed through additional photometry with the Transiting Exoplanet Survey Satellite (TESS). Other ground-based follow-up observations were carried out at the South African Astronomical Observatory (SAAO) with the 1.0 m and 1.9\,m telescopes, as well as with EulerCam on the 1.2m Euler Telescope at La Silla Observatory. Radial velocity measurements made with the HARPS, CORALIE and FEROS spectrographs allowed for the masses of all four planets to be determined. \

Spectral analysis via \textsc{iSpec} and SED fitting via \textsc{ARIADNE} revealed the properties of the host stars in each system, which were found to be F, G or K-type stars (see Table \ref{tbl:Spectral Properties}). We also attempted to constrain the rotation periods of the stars and from this determined that the four stars were likely to be slowly rotating. Global fits to the data were produced using the open-source astronomy software package \textsc{allesfitter}. Our results for HATS-54b (NGTS-22b) are consistent with the original discovery paper \citep{Espinoza2019}. From the orbital decay time-scales of the 4 planets presented in this paper, they all appear to be on stable orbits while the host star remains on the main-sequence. \

All four planets receive a level of irradiation above the minimum threshold for the onset of planetary inflation mechanisms \citep{Demory2011,Miller2011}. Therefore, we sought to determine whether any of the planets experienced inflation by comparing our derived radii with predictions from evolutionary models \citep{Baraffe2008}. In addition, we examined the predicted additional radius change due to inflation, $\Delta R$, through the flux-mass-radius relations outlined in \citet{Sestovic2018}. We found that NGTS-23b is likely inflated while NGTS-24b is more likely to be inflated than not inflated. NGTS-25b is consistent with both inflated and non-inflated models so we are unable to determine whether it is inflated. HATS-54b is unlikely to be inflated as it is  consistent with the non-inflated models. \

Finally, similar to \citet{Tilbrook2020}, we note that there is a disparity between some of our derived radii and those predicted by the inflationary forward model of \citet{Sestovic2018}. Through the inclusion of both new hot Jupiter data and additional parameters, such as system age and heavy metal fraction, the accuracy of inflationary parameterisation can be improved, which would help improve our understanding of the physical processes behind the inflation of hot Jupiters.

\section*{Acknowledgements}

Based on data collected under the NGTS project at the ESO La Silla Paranal Observatory. 
The NGTS facility is operated by the consortium institutes with support from the UK Science and Technology Facilities Council (STFC) ST/M001962/1, ST/S002642/1 and ST/W003163/1.
This study is based on observations collected at the European Southern Observatory under ESO programmes 0102.A-9011, 0103.A-9004, 0103.C-0719, 0104.A-9012, 0104.C-0588 and 105.20G9.
We acknowledge the use of public TESS data from pipelines at the TESS Science Office and at the TESS Science Processing Operations Centre.
This paper includes data collected with the TESS mission, obtained from the MAST data archive at the Space Telescope Science Institute (STScI). Funding for the TESS mission is provided by the NASA Explorer Program. STScI is operated by the Association of Universities for Research in Astronomy, Inc., under NASA contract NAS 5–26555.
This paper uses observations made at the South African Astronomical Observatory (SAAO).
JSJ greatfully acknowledges support by FONDECYT grant 1201371 and from the ANID BASAL projects ACE210002 and FB210003.
JIV acknowledges support of CONICYT-PFCHA/Doctorado Nacional-21191829.
The contributions at the University of Warwick by PJW, RGW and SG have been supported by STFC through consolidated grants ST/L000733/1 and ST/P000495/1. DJA acknowledges support from the STFC via an Ernest Rutherford Fellowship (ST/R00384X/1).
MNG acknowledges support from the European Space Agency (ESA) as an ESA Research Fellow.
JAGJ acknowledges support from grants HST-GO-15955.004-A and HST-AR-16617.001-A from the Space Telescope Science Institute, which is operated by the Association of Universities for Research in Astronomy, Inc., under NASA contract NAS 5-26555. JAGJ acknowledges support from NASA under grant number \textcolor{black}{80NSSC22K0125} from the TESS Cycle 4 Guest Investigator Program.
The contributions of FB, ML, OT, and SU have been carried out within the framework of the NCCR PlanetS supported by the Swiss National Science Foundation under grants 51NF40\_182901 and 51NF40\_205606. ML also acknowledges support of the Swiss National Science Foundation under grant number PCEFP2\_194576.
DGJ acknowledges the support of the Department for Economy (DfE).

\section*{Data Availability}

The data underlying this article will be shared on reasonable request
to the corresponding author.



\bibliographystyle{mnras}
\bibliography{refs} 



\appendix

\section{Additional Photometry}

\begin{figure*}
    \centering
    \begin{subfigure}[b]{0.475\textwidth}
        \centering
        \includegraphics[width=\textwidth]{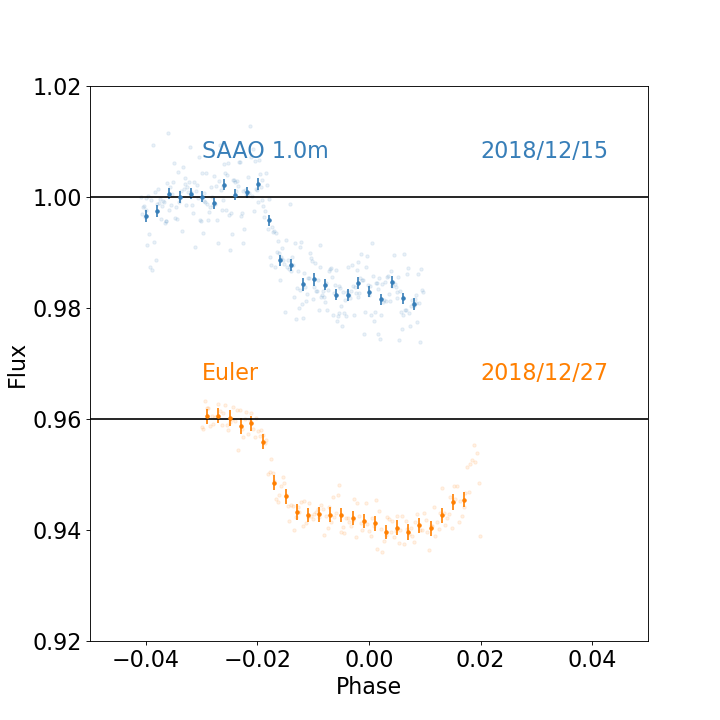}
        \caption[]%
        {{\small Additional Photometry for NGTS-23b.}}    
        \label{fig:appendix_HATS54b}
    \end{subfigure}
    \hfill
    \begin{subfigure}[b]{0.475\textwidth}   
        \centering 
        \includegraphics[width=\textwidth]{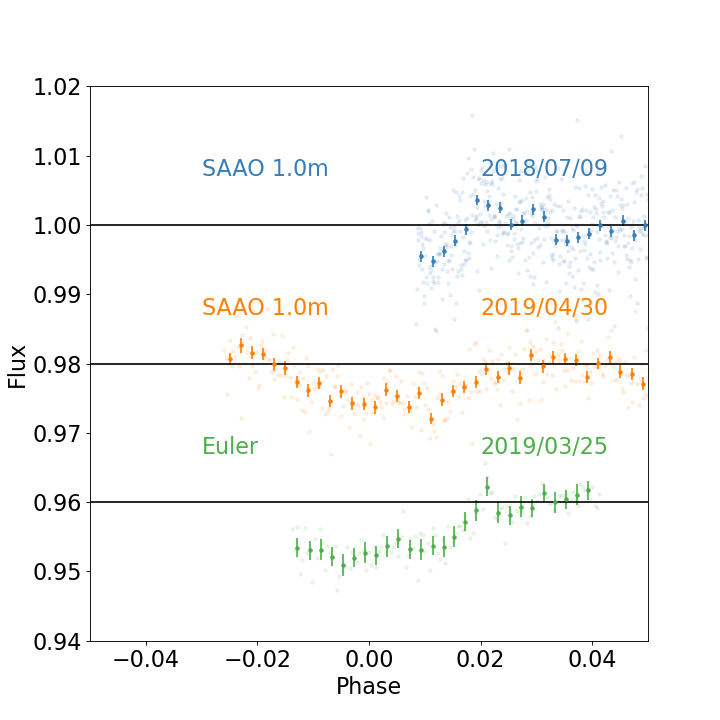}
        \caption[]%
        {{\small Additional Photometry for HATS-54b (NGTS-22b). }  }
        \label{fig:appendix NGTS-23b}
    \end{subfigure}
     \caption[]
    {\small Phase-folded photometry for NGTS-23b and HATS-54b (NGTS-22b). These transits were not used in the fitting procedure as there was not sufficient baseline to characterise the out of transit behaviour.} 
    \label{fig:appendix phot plots}
\end{figure*}

\section{Sample NGTS photometry}

\begin{table*}
    \caption[]
    {\small Sample NGTS photometry for each object. The full tables are available in a machine-readable format from the online journal.} 
    \centering
    \begin{subtable}[b]{0.475\textwidth}
        \centering
        \begin{tabular}{c c c }
        \hline
        Time & Normalised Flux & Flux error\\  
        (BJD-2450000) & & \\
        \hline
        7981.8851 & 1.016 & 0.022\\
        7981.8852 & 1.002 & 0.022\\
        7981.8854 & 0.978 & 0.022\\
        7981.8855 & 0.999 & 0.022\\
        7981.8857 & 1.022 & 0.022\\
        ... & ... & ... \\
        8195.5415 & 1.020 & 0.020\\
        8195.5416 & 1.007 & 0.020\\
        8195.5417 & 1.056 & 0.020\\
        8195.5419 & 1.012 & 0.020\\
        8195.5420 & 0.956 & 0.020\\
        \hline
        \end{tabular}
        \caption[]%
        {{\small NGTS Photometry for NGTS-23.}}    

    \end{subtable}
    \hfill
    \begin{subtable}[b]{0.475\textwidth}  
        \centering
        \begin{tabular}{c c c }
        \hline
        Time & Normalised Flux & Flux error\\  
        (BJD-2450000) & & \\
        \hline
        8097.8107 & 0.993 & 0.010\\
        8097.8108 & 0.985 & 0.010\\
        8097.8110 & 0.977 & 0.010\\
        8097.8111 & 1.011 & 0.010\\
        8097.8113 & 1.013 & 0.010\\
        ... & ... & ... \\
        8295.5989 & 0.969 & 0.014\\
        8295.5991 & 1.001 & 0.014\\
        8295.5992 & 0.975 & 0.014\\
        8295.5993 & 1.013 & 0.014\\
        8295.5995 & 1.007 & 0.014\\
        \hline
        \end{tabular}
        
        \caption[]%
        {{\small NGTS Photometry for NGTS-24.}}    

    \end{subtable}
    \vskip\baselineskip
        \begin{subtable}[b]{0.475\textwidth}   
        \centering
        \begin{tabular}{c c c }
        \hline
        Time & Normalised Flux & Flux error\\  
        (BJD-2450000) & & \\
        \hline
        8203.8697 & 1.009 & 0.022\\
        8203.8698 & 0.984 & 0.022\\
        8203.8700 & 1.051 & 0.022\\
        8203.8701 & 0.983 & 0.022\\
        8203.8702 & 0.959 & 0.022\\
        ... & ... & ... \\
        8423.5825 & 1.001 & 0.022\\
        8423.5826 & 1.020 & 0.022\\
        8423.5828 & 0.968 & 0.022\\
        8423.5829 & 0.954 & 0.022\\
        8423.5831 & 0.984 & 0.022\\
        \hline
        \end{tabular}

        \caption[]%
        {{\small NGTS Photometry for NGTS-25.}}    

    \end{subtable}
    \hfill
    \begin{subtable}[b]{0.475\textwidth}   
        \centering
        \begin{tabular}{c c c }
        \hline
        Time & Normalised Flux & Flux error\\  
        (BJD-2450000) & & \\
        \hline
        7374.8169 & 0.991 & 0.017\\
        7374.8171 & 0.999 & 0.017\\
        7375.8065 & 0.979 & 0.019\\
        7375.8068 & 0.970 & 0.019\\
        7375.8069 & 1.004 & 0.019\\
        ... & ... & ... \\
        7632.4995 & 0.996 & 0.018\\
        7632.4997 & 0.994 & 0.018\\
        7632.4998 & 0.982 & 0.018\\
        7632.5000 & 0.968 & 0.018\\
        7632.5001 & 1.018 & 0.018\\
        \hline
        \end{tabular} 

        \caption[]%
        {{\small NGTS Photometry for HATS-54 (NGTS-22).}}    

    \end{subtable}

    \label{tbl:NGTS sample photometry}
\end{table*}

\section{Radial Velocity Data}

\begin{table*}
    \centering
    \caption{Radial velocity data for the four systems.}
    \begin{tabular}{c c c c c c c c}
    \hline
    Target & BJD & RV & RV error & FWHM & contrast & BIS & instrument\\
    & (-2450000) & ($\rm{km\,s^{-1}}$) & ($\rm{km\,s^{-1}}$) & ($\rm{km\,s^{-1}}$) & &  ($\rm{km\,s^{-1}}$) & \\
    \hline\hline
    
    \multirow{18}{*}{NGTS-23}
    &8780.748 & 22.344 & 0.058 & 8.499 & 29.492 & 0.085 & HARPS\\
    &8781.748 & 22.325 & 0.033 & 8.376 & 29.162 & 0.016 & HARPS\\
    &8783.744 & 22.217 & 0.033 & 8.406 & 28.960 & 0.024 & HARPS\\
    &8784.756 & 22.315 & 0.040 & 8.445 & 29.733 & 0.162 & HARPS\\
    &8843.670 & 22.306 & 0.034 & 8.477 & 28.427 & 0.198 & HARPS\\
    &8844.657 & 22.180 & 0.023 & 8.376 & 28.400 & 0.104 & HARPS\\
    &8868.586 & 22.212 & 0.018 & 8.374 & 27.577 & 0.017 & HARPS\\
    
    &8418.650 &	22.415 & 0.080 & 8.946 & 31.387 & -0.019 & CORALIE\\	
    &8424.713 &	22.254 & 0.076 & 8.926 & 31.042 & 0.703 & CORALIE \\	
    &8428.803 &	22.292 & 0.075 & 9.093 & 31.454 & -0.112 & CORALIE \\	
    
    &8736.888 & 22.281 & 0.028 & - & - & 0.082 & FEROS	\\
    &8798.817 & 22.176 & 0.032 & - & - & -0.101 & FEROS \\
    &8800.690 & 22.245 & 0.025 & - & - & 0.025 & FEROS \\
    &8847.709 & 22.288 & 0.023 & - & - & 0.106 & FEROS \\
    &8848.621 & 22.216 & 0.023 & - & - & -0.009 & FEROS \\
    &8850.645 & 22.390 & 0.019 & - & - & -0.014 & FEROS \\
    &8851.688 & 22.291 & 0.022 & - & - & 0.050 & FEROS \\
    &8852.642 & 22.199 & 0.019 & - & - & 0.111 & FEROS \\
    \hline
    
    \multirow{11}{*}{NGTS-24} 
    &8595.671 & -15.272 & 0.084 & 9.264 & 42.373 & -0.011 & CORALIE\\	
    &8640.590 & -15.180 & 0.084 & 9.176 & 42.392 & -0.088 & CORALIE\\	
    &8667.515 &	-15.296 & 0.084 & 9.022 & 43.598 & 0.109 & CORALIE\\
    &8829.810 & -15.364 & 0.074 & 9.026 & 37.061 & -0.004 & CORALIE\\	
    &8843.768 & -15.394 & 0.085 & 9.113 & 43.453 & 0.237 & CORALIE\\
    
    &8847.794 & -15.335 & 0.011 & - & - & 0.014 & FEROS \\	
    &8848.802 & -15.247 & 0.012 & - & - & 0.026	& FEROS \\
    &8849.777 & -15.305 & 0.014 & - & - & -0.011 & FEROS \\
    &8850.770 & -15.347 & 0.011 & - & - & 0.027 & FEROS \\
    &8851.764 & -15.229 & 0.012 & - & - & 0.034 & FEROS \\
    &8852.769 & -15.270 & 0.011 & - & - & -0.007 & FEROS \\
    \hline
    
    \multirow{14}{*}{NGTS-25}
    &9377.789 & -22.926	& 0.009 & 6.801 & 41.105 & 0.057 & HARPS \\
    &9411.722 & -22.913 & 0.011 & 6.878 & 40.947 & 0.054 & HARPS \\
    &9412.775 & -23.060 & 0.014 & 6.819 & 41.233 & 0.073 & HARPS \\
    &9426.658 & -23.009 & 0.021 & 6.841 & 40.176 & 0.007 & HARPS \\	&9428.610 & -22.931 & 0.015 & 6.897 & 40.651 & 0.011 & HARPS \\	&9429.675 & -23.068 & 0.014 & 6.742 & 41.160 & -0.027 & HARPS \\	&9430.706 & -23.112 & 0.056 & 6.927 & 37.081 & 0.117 & HARPS \\	&9432.638 & -23.093 & 0.014 & 6.934 & 40.522 & 0.007 & HARPS \\	&9433.636 & -23.032 & 0.016 & 6.829 & 40.429 & 0.068 & HARPS \\		&9461.668 & -23.055 & 0.017 & 6.748 & 41.118 & 0.073 & HARPS \\	&9462.671 & -22.923 & 0.017 & 6.761 & 40.849 & 0.044 & HARPS \\	&9470.521 & -22.987 & 0.019 & 6.845 & 40.277 & 0.089 & HARPS \\
    
    &8704.642 & -23.285 & 0.165 & 8.407 & 92.508 & 0.168 & CORALIE \\
    &8717.567 & -22.566 & 0.227 & 8.508 & 193.087 & 0.105 & CORALIE \\
    \hline
    
    \multirow{19}{5em}{HATS-54 (NGTS-22)}
    &8613.692 & 46.189 & 0.011 & 7.858 & 50.059 & -0.064 & HARPS \\	&8615.664 & 46.066 & 0.007 & 7.837 & 50.219 & -0.031 & HARPS \\	
    &8616.656 & 46.191 & 0.008 & 7.837 & 49.736 & -0.003 & HARPS \\	&8617.649 & 45.981 & 0.014 & 7.847 & 48.371 & -0.010 & HARPS \\		&8642.655 & 46.064 & 0.009 & 7.798 & 50.454 & 0.033 & HARPS \\

    &8546.699 & 46.124 & 0.151 & 9.206 & 58.315 & -4.184 & CORALIE \\		&8547.824 & 46.269 & 0.099 & 8.644 & 55.326 & -0.176 & CORALIE \\	&8565.688 & 46.198 & 0.118 & 8.889 & 49.903 & 0.080 & CORALIE \\	&8566.837 & 45.971 & 0.084 & 8.581 & 51.157 & -0.218 & CORALIE \\	
    &8593.533 & 46.200 & 0.132 & 8.649 & 49.514 & 0.141 & CORALIE \\	&8601.602 & 46.208 & 0.094 & 8.737 & 53.746 & -0.085 & CORALIE \\	
    
    &8591.597 & 46.100 & 0.011 & - & - & -0.022 & FEROS \\
    &8592.680 & 46.135 & 0.029 & - & - & 0.004 & FEROS \\
    &8593.569 &	46.227 & 0.013 & - & - & 0.005 & FEROS \\
    &8594.619 &	45.938 & 0.013 & - & - & 0.006 & FEROS \\
    &8597.574 & 45.965 & 0.014 & - & - & -0.075 & FEROS \\
    &8599.601 & 46.040 & 0.013 & - & - & -0.037 & FEROS \\
    &8599.648 & 45.968 & 0.014 & - & - & 0.006 & FEROS \\
    &8599.798 & 46.007 & 0.015 & - & - & -0.066 & FEROS \\
    \hline
    \end{tabular}
\end{table*}

\section{NGTS periodograms}

\begin{figure*}
    \centering
    \begin{subfigure}[b]{0.475\textwidth}
        \centering
        \includegraphics[width=\textwidth]{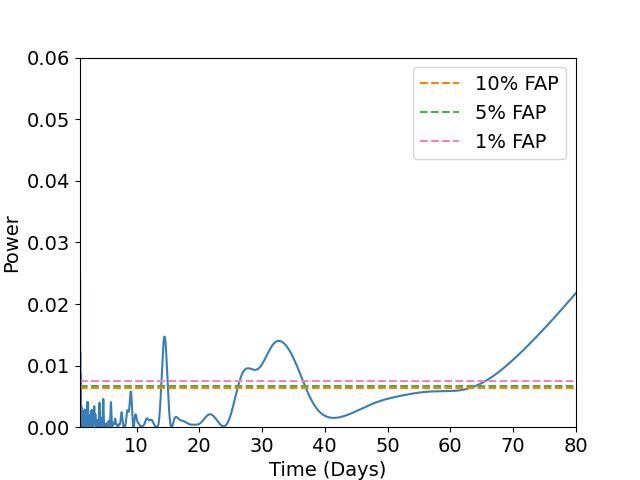}
        \caption[]%
        {{\small NGTS-23b}}    

    \end{subfigure}
    \hfill
    \begin{subfigure}[b]{0.475\textwidth}  
        \centering 
        \includegraphics[width=\textwidth]{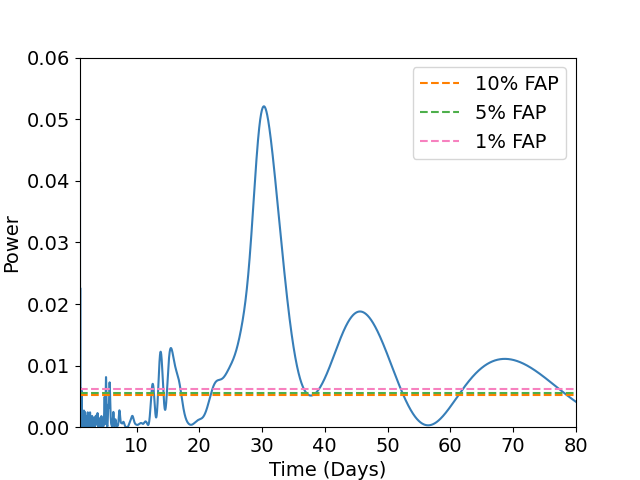}
        \caption[]%
        {{\small NGTS-24b}}    

    \end{subfigure}
    \vskip\baselineskip
    \begin{subfigure}[b]{0.475\textwidth}   
        \centering 
        \includegraphics[width=\textwidth]{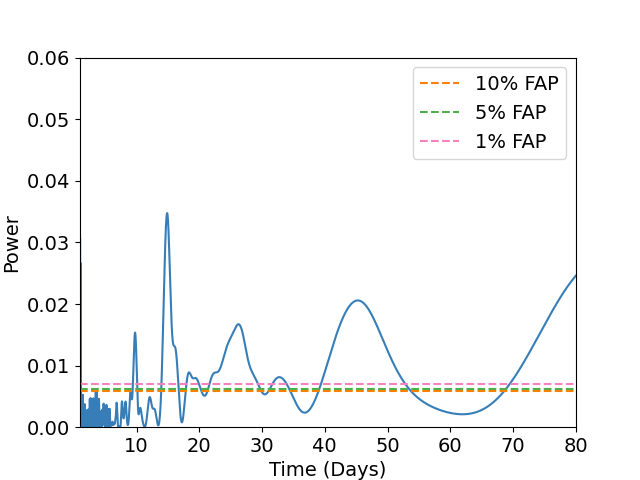}
        \caption[]%
        {{\small NGTS-25b}}    

    \end{subfigure}
    \hfill
    \begin{subfigure}[b]{0.475\textwidth}   
        \centering 
        \includegraphics[width=\textwidth]{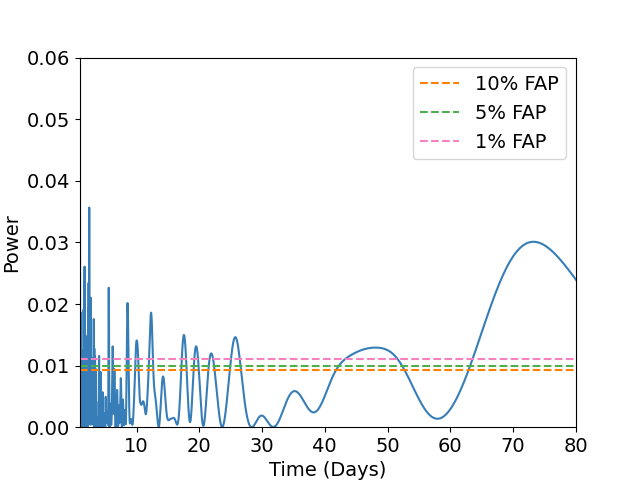}
        \caption[]%
        {{\small HATS-54b (NGTS-22b)}}    

    \end{subfigure}

    \caption[]
    {\small GLS periodograms for the NGTS photometry.} 
    \label{fig:NGTS periodograms}
\end{figure*}


\bsp	
\label{lastpage}
\end{document}